\documentclass[12pt,preprint]{aastex}

\newcommand{\kms}{$\;$km s$^{-1}$}
\newcommand{\msun}{M_{\odot}}
\newcommand{\rsun}{R_{\odot}}
 
\begin{document}
 
\title{The Blue Straggler RS CVn Star S1082 in M67:
A Detailed Light Curve and the Possibility of a Triple\footnote{ Based in 
part on observations obtained with the
Hobby-Eberly Telescope, which is a joint project of the University of
Texas at Austin, the Pennsylvania State University, Stanford
University, Ludwig-Maximillians-Universit\"at M\"unchen, and
Georg-August-Universit\"at G\"ottingen.} \footnote{ Based in part on 
observations obtained with the Multiple Mirror Telescope,
a joint facility of the Smithsonian Institution and the University of
Arizona.}}

\author{Eric L. Sandquist}
\affil{San Diego State University,Department of Astronomy,San Diego, CA 92182}
\email{erics@mintaka.sdsu.edu}

\author{David W. Latham} 
\affil{Harvard-Smithsonian Center for Astrophysics, 60 Garden Street,
Cambridge, MA 02138} 
\email{dlatham@cfa.harvard.edu}

\author{Matthew D. Shetrone}
\affil{University of Texas/McDonald Observatory,
P.O. Box 1337, Fort Davis, Texas 79734}
\email{shetrone@astro.as.utexas.edu}

\author{Alejandra A. E. Milone}
\affil{MMT Observatory, P.O. Box 97, Amado, Arizona, 85645}
\email{amilone@mmto.org}
\begin{abstract}

Using both photometric and spectroscopic data, we present a picture of
the very unusual blue straggler S1082 in the old open cluster M67
whose light is the sum of a close binary ($P = 1.0677978$ d) and 
another cluster member. The primary of the close binary and the third star
are both blue stragglers in their own rights. Using
relative photometry with millimagnitude accuracy we provide a complete
$V$-band light curve for the system, and show a number of unusual
features: brightness variations at the 0.01-0.03 mag level from month
to month at all phases, a narrow primary and broad secondary eclipse,
brightness differences between phases 0.25 and 0.75, and short
duration ($\sim 1$ h) drops in brightness. Much of the light curve
variation appears to be due to spot activity on the cooler, fainter,
nearly synchronized component of the close binary. We use spectra from
several sources to constrain the temperatures of the three known
components, the relative flux contributions, rotational velocities,
and radial velocities. The data clearly show that the brightest star
(narrow-line component) seen in the spectrum is on an orbit with
period $P = 1189 \pm 7$ d and eccentricity $e = 0.57 \pm 0.08$,
although we cannot prove that there is a dynamical link between it and
the close binary. The systematic velocities of all of the stars
indicate that they are members of the cluster.  Our models of the
system indicate that the cooler component of the close binary lies on
the main sequence near the cluster turn-off, while the hotter
component lies near an extension of the main sequence blueward of the
turn-off, and thus qualifies as a blue straggler by itself.  The
reduction of the masses of the stars in the close binary compared to
previous models indicates that it is possible that the more
massive component of the close binary formed from a merger of just two
turnoff-mass stars.

\end{abstract}


\section{Introduction}

In the past two decades, radial-velocity surveys and time-series
photometry studies have begun to provide strong clues to the
mechanisms that form blue stragglers: stars brighter and bluer than
the turnoff for their stellar population.  Mass transfer in binaries
and dynamical interactions between binaries have emerged as the two
leading formation mechanisms (e.g. Leonard 1996). For clusters like
the old open cluster M67, the large sample of extensively observed
blue stragglers presents the opportunity of understanding the relative
frequencies at which the different formation mechanisms produce
stragglers.

Radial-velocity studies of the classical blue stragglers (Milone \&
Latham 1994) have proven to be especially illuminating.  Among the 10
blue stragglers that are rotating slowly enough to yield reliable
velocities with the CfA Digital Speedometers, 6 are single-lined
spectroscopic binaries for which preliminary orbits have been reported
(Latham \& Milone 1996).  The closest binary (S1284, designation from Sanders 1977; also known as F190, designation from Fagerholm 1906) has a
period, $P = 4.18258 \pm 0.00006$ d, and mass function, $f(M) = 0.0017
\pm 0.0013$ (Milone \& Latham 1992; the values reported here are based
on an updated orbital solution), that are consistent with an
evolutionary history involving stable mass transfer from an initially
more massive companion that is now a white-dwarf remnant (e.g.,
Rappaport et al.\ 1995).  However, the orbit has a modest but
significant eccentricity, $e = 0.256 \pm 0.034$, which is hard to
understand.  Neither stable mass transfer nor weak gravitational
encounters with other cluster members can easily create such a large
eccentricity.  The other 5 blue stragglers with CfA orbits all have
relatively long periods, ranging from 850 to 5153 days (Latham \&
Milone 1996; the values reported here are also based on updated
orbital solutions). Two have much more eccentric orbits than is
expected from stable mass transfer (S1267: $P = 850 \pm 11$ d, $e =
0.47 \pm 0.11$; S997: $P = 5153 \pm 87$ d, $e = 0.357 \pm 0.039$).
This leaves dynamical interactions between binaries as the leading
contender for the formation mechanism responsible for these two blue
stragglers (e.g., Leonard 1996, Hurley et al.\ 2001).  The other three
have lower orbital eccentricities, all within $3\sigma$ of circular
(S752: $P=1013 \pm 10$ d, $e = 0.27 \pm 0.10$; S1195: $P = 1139 \pm
12$ d, $e = 0.025 \pm 0.075$; S975: $P = 1231 \pm 8$ d; $e = 0.124 \pm
0.048$).  Stable mass transfer remains the leading contender for the
formation mechanism responsible for these blue stragglers because it
is hard to end up with nearly circular orbits from strong multiple
star interactions.

By examining blue stragglers in the field, it is possible to more
cleanly examine the effects that mass transfer has on a binary by
eliminating from consideration stragglers that formed during binary
interactions in a dense stellar environment. Carney et al.\ (2001)
studied a sample of 10 isolated metal-poor field blue stragglers, six
of which proved to be single-lined spectroscopic binaries, again with
relatively long periods, ranging from 167 to 844 days, and low
eccentricities.  The mass functions are consistent with all the unseen
companions being white dwarf remnants, and Carney et al. (2001) argued
that stable mass transfer was the mechanism responsible for the
formation of these blue stragglers.  A similar conclusion was reached
by Preston \& Sneden (2000) for a sample of blue metal-poor stars that
they studied.

Coalescence of two stars in a close binary undergoing unstable mass
transfer and common envelope evolution is undoubtedly a potential
source of relatively massive stars with extended main-sequence
lifetimes, some of which may be massive enough to move blueward of the
turn-off to become blue stragglers by the classical definition. For
example, S1036 is known to be a W UMa contact binary that is doomed to
end in a merger (e.g. Mateo et al. 1990), and two other W UMa systems,
S757 and S1282 are located near the turn-off in M67.  However, the end
products of mass transfer and stellar mergers do not necessarily have
to appear brighter and bluer than the turn-off.  Indeed, the
color-magnitude diagram of M67 is sprinkled with a number of stars
that are kinematic members of the cluster but fall well to one side or
the other of the main locus of stars on the diagram.  Good examples
are S1040, S1072, S1113, and S1063 (e.g., Landsman et al. 1997;
Mathieu et al.\ 2002).  Mass transfer and/or mergers have been
explored as possible explanations for these stars, with limited
success (e.g., Hurley et al.\ 2001, Mathieu et al.\ 2002).

S1082 (ES Cnc; F131; MMJ6493, designation from Montgomery, Marschall,
\& Janes 1993) has long been identified as a blue straggler in M67,
and has consistently been shown to be a member of the cluster by
proper motion studies \citep{sanders77,girard89,zhao93}. In the last
13 years the system has come under close scrutiny.  A broad secondary
component was first detected in the spectrum by Mathys (1991) and
Prichet \& Glaspey (1991).  In the same year, Simoda (1991) reported
the system to be a photometric variable. Using 10 years of photometric
data, Goranskij et al. (1992) discovered S1082 to have a close binary
with partial eclipses and a period of $1.0677978 \pm 0.0000050$ d.
Oddly, the radial-velocity studies of Manteiga et al. (1989) and
Milone \& Latham (1994) found some velocity variability but no sign of
a period consistent with the photometric period. S1082 is one of just
three blue stragglers in open clusters that are known to be X-ray
sources (Belloni \& Verbunt 1996; Belloni \& Tagliaferri 1998). A
ROSAT survey of the field (Belloni, Verbunt, \& Schmitt 1993) found
S1082 to be one of the brighter sources in M67. Based on a measured UV
excess, Landsman et al. (1998) hypothesized that the system was
a binary with a hot subluminous companion.  van den Berg et al. (1999)
confirmed the radial-velocity variability of the broad component first
seen by Mathys (1991) in the O triplet, the Na D lines and H$\alpha$.
Shetrone \& Sandquist (2000) uncovered a larger number of lines for
the broad component and found that the velocity amplitude was large
and had a short period, while the narrow component did not seem to
show any reflex motion from the broad component.  In a recent paper,
van den Berg et al. (2001; hereafter vOVS) presented photometric and
spectroscopic data on S1082 in which they discovered a third stellar
component in the spectrum, and proposed that the system is a
hierarchical triple with two of the stars being blue stragglers in
their own rights [one of the stars in the inner binary (labelled
component Aa by vOVS), and the outermost star (component B)]. In this
paper, we present observations that were taken independently of those
of vOVS, and which in many ways complement theirs.

vOVS also stated that S1082 is likely to be a RS CVn system. Photometric
studies of many RS CVn systems find that starspots are almost
universally present, and their presence can affect derived system
parameters unless carefully modeled (Hilditch \& Bell 1994). The
relatively small number of observations made by vOVS made spot
modeling impractical.  In \S 2 of this paper we present a
considerably larger body of photometric observations in an attempt to
determine a ``normal'' light curve for the system and to estimate the
degree to which the presence of spots affects the light curve and
binary parameters derived from it. In \S 3, we present spectroscopic
observations that constrain the nature of the orbit of the brightest
star in S1082 (component B in vOVS) and that provide independent
measurements of the temperatures of the components.  In \S 4 we
discuss the nature of the system.
 
\section{Photometric Observations}

The light curves that have been presented for S1082 have necessarily
been somewhat fragmented due to the near 1 d period of the close
binary.  Although the period of the binary is known very well (compare
the Goranskij et al. period with the revised period from vOVS:
$1.0677971 \pm 0.0000007$ d), a complete light curve taken within a
small number of orbits has not yet been presented. The Goranskij et
al. light curve was incomplete (it didn't completely cover the
secondary eclipse), and by their own admission the photometry from the
different telescopes used could not be properly put on a common
photometric system. vOVS provided data for the secondary eclipse in
the light curve, although their light curve is not complete through
the primary eclipse.

All of the photometry for this study was taken at the 1 m telescope at
the Mt. Laguna Observatory using a $2048 \times 2048$ CCD on nights
between December 2000 and March 2002. The nights of observations are
given in Table~\ref{obs}. The photometry was in the $V$ band
with typical exposure times of 20 s (ranging between 15 and 60 s
depending on atmospheric conditions) to optimize the counts for S
1082. Exposures were usually separated by about 2.5 minutes due to a
relatively long readout time for the CCD. This dataset covers the
entire binary period with several observations of most
phases. Our photometry was taken one and two years after the
photometry of vOVS, and so will also be useful in gauging variability
timescales for the system.

Guiding jitter in the telescope typically restricted image quality to
greater than 4 pixels (1.6 arcsec) in the best seeing conditions.
Observing conditions for the nights varied greatly, reaching FWHM of
13 pixels for some of the worst frames. The relative sparseness of the
cluster worked in our favor in this case though, because only pairs of
stars with the smallest separations were affected.  In addition,
several hundred columns of the chip had charge transfer problems
during the December 2000 run. In this case measurements of stars that
fell near these columns were eliminated. The remainder of the chip
does not seem to have been noticeably affected by this problem. In the
January and February 2001 runs this problem was almost entirely
corrected. The CCD was replaced for the March 2001 run and a
two-amplifier readout was employed, which halved the duty cycle for
our observations.

The object frames were processed in usual fashion, using overscan
subtraction, bias frames, and flat fields (usually twilight flats,
with the exception of nights d7, f1, and f2, for which dome flats had
to be used; see Table~\ref{obs} for explanation of the night
identifications).  We chose to rely on aperture photometry for this
study, using the IRAF\footnote{IRAF (Image Reduction and Analysis
Facility) is distributed by the National Optical Astronomy
Observatories, which are operated by the Association of Universities
for Research in Astronomy, Inc., under contract with the National
Science Foundation.} tasks DAOFIND and PHOT from the APPHOT
package. Crowding sometimes prevented measurements in the largest
apertures, however. Curve-of-growth analysis was conducted using the
IRAF tool DIGIPHOT.PHOTCAL.MKAPFILE in order to bring all photometric
measurements to the same total aperture size. The general procedure is
discussed in Stetson (1990).  Individual nights were run separately
through the curve-of-growth analysis.

In order to improve the accuracy of the relative photometry for the
light curves, we used an ensemble photometry method similar to that
described by Honeycutt (1992). The method described by Honeycutt is
essentially a simultaneous least-squares solution for mean magnitudes
of all stars observed and the relative zero points of all image frames
used. This method allows for the reduction of
observations of distinct but mutually overlapping fields to a
consistent photometric system. Honeycutt's algorithm was used to
obtain a preliminary solution for all of our photometry. Least-squares
methods tend to be overly-sensitive to outlier points, however. On
nights with poor seeing it was still possible to obtain good aperture
photometry for the majority of the stars on each frame, but image
blending would contaminate some measurements. These frames can still
be accurately brought to the common photometric system, but care
must be taken to deal with poor measurements. We opted to eliminate
stars with less than 20 measurements, and with large scatter around
the mean.  However, the most useful improvement was in using medians
to determine the final zero points and magnitudes. This was done in an
iterative manner, using the least-squares solution for mean magnitudes
to determine the median zero-point value, and then using the
redetermined zero points to determine median magnitudes for each
star. Iteration continued until the solution converged to better than
0.0005 mag for both the frame zero points and star magnitudes. 

We also allowed for the possibility that magnitude residuals could be
a function of star position on the CCD for each frame
\citep{gill88}. Second order polynomials were fit to the residuals in
the x- and y-directions, and subtracted from individual
measurements. The polynomials were well-determined because of the
large numbers of stars measured on each frame. The significance of the
position-dependent correction varied, but at most amounted to a few
times 0.01 mag. The corrections were important in establishing the zero points
because the frame center did drift relative to the sky, and because
the position of the frame center was sometimes changed by nearly the width of
the frame from night-to-night.

We found that this procedure was robust in that it did not require us
to manually prune the data. We could also leave the variable stars in
the solution with no ill effects --- the variables (particularly
low-amplitude ones) do contain usable information that constrains the
frame zero points.  Because frames typically had more than 300
measurable stars, the formal errors in the zero points ranged from
around 0.002 to 0.005 mag, even with respect to night-to-night
variations. Typical errors in individual measurements of S1082 were
between 0.003 and 0.007 mag. This level of accuracy was important for
detecting features in the light curves of several of the variable
stars.

\subsection{The Light Curve}\label{analysis}

The phased data for all observations of S1082 are shown in
Fig.~\ref{allphot}. Two partial eclipses of slightly
different depths (but seemingly different widths) are clearly
present. To make the variability of the light curves clearer, we plot
the data from each month separately in Fig.~\ref{monthphot}. Data from
successive nights with the most overlap (d7/d8, j1/j3, j7/j8/j9) show
that variability in the light curve on the order of days is minimal,
and the smoothness of light curves from successive nights without much
overlap (d1/d3, f1/f2, m1/m3) supports this. Data from successive
months clearly do show variations in the light curves.

Much of the short timescale variability in the light curve is probably
real, but currently beyond our ability to model adequately. Variations
in many portions of the light curve amount to up to 0.03 mag from
month to month. If this is due to coverage by starspots (in accordance
with the idea that it is an RS CVn variable), then the binary may have
strong magnetic activity. Variations in the light curve of XY UMa, perhaps the
most active member of the class, can be as large as 0.2 mag (Pribulla
et al. 2001). The analysis of S1082 is complicated by the large
``third light'' component (which will be discussed in \S
\ref{tsecm}). In comparison to the total amplitude of the light curve
(0.115 mag) of S1082 the variability of the light curve is a serious
complication. Before turning to modeling, we discuss general features
of the light curve below.

\subsubsection{Primary Eclipse}

The primary eclipse is partial, with a total duration that is 13\% of
the orbit. (Our observations on night m1.2 are the only ones that
cover this eclipse in its entirety.) The depth of primary eclipse
appears to vary by at least 0.015 mag. The maximum depth may
occasionally be larger than was seen on nights d1 and m3, but we have
only two observations to support this (with an observation on night m5
being the most reliable). There are hints of this variability in the
photometry of Goranskij et al., but the interpretation of their data
is complicated by their inability to remove zero-point differences in
data taken on different telescopes. The single eclipse that vOVS
observed was on the shallower side of the range we observed.

\subsubsection{Secondary Eclipse}

Our most complete observations of the secondary eclipse occurred
during January 2001 and February 2002. In both cases, the light
variation was continuous for very nearly 50\% of the phase, with no
obvious indication of sharper change near the expected contact
points. On three occasions (nights d3, j1, and f1.2), the light curve
rose almost linearly up to phase 0.75, and immediately turned downward
after. Similar features are seen near phase 0.25 on several nights,
but there is larger scatter in the observations of that quadrature
phase. vOVS noted that their observations of the secondary eclipse
(like our observations on nights j1 and j3) showed an unusually
constant slope to the light curve from near secondary minimum to near
phase 0.75.

A comparison of the light curves in Fig.~\ref{allphot} clearly shows
that the shape and depth of the secondary eclipse is quite variable
though --- features seen at phases 0.4 and 0.56 on night d8 might be
indications of eclipse contacts, but these are the only relatively
clear signs. These slope breaks appear to fall at approximately the
same brightness level (approximately 0.025 mag below peak light) as
seen for the primary eclipse. The eclipses
observed on several nights are asymmetric.

The observations indicate that the depth of the secondary eclipse is
variable, but to about the same degree (about 0.02 mag) as the primary
eclipse for the time period of our observations. The variations in
shape are reminiscent of variations seen in the active RS CVn variable
XY UMa (Collier Cameron \& Hilditch 1997; Pribulla et al. 2001), which
is inferred to have starspots. That system showed similar behavior
during secondary eclipses, indicating spotting near a longitude of
$0\degr$. The several month timescale for the light curve variability
is also consistent with our observations. This topic is important
because it significantly affects the modeling of the light curve.

\subsubsection{Quadratures}

Goranskij et al. indicated that the secondary maximum at phase 0.75
was brighter than the primary maximum at phase 0.25. vOVS verified
this, but had large scatter in their data around primary maximum. In
our data, there does not seem to be noticeable differences in the
brightness level, although there is clear variability in the primary
maximum of up to 0.03 mag from run to run.  vOVS also indicate that
they found a large amount of scatter in their measurements between
phases 0.1 and 0.3, although they attributed it to poor observing
conditions.

A feature of the secondary quadrature that we have been unable to
model adequately was the sharp change in the slope of the light curve
near phase 0.75 in several of the nights of data (d3, j1, f1.2). This
is inconsistent with any sort of ellipsoidal variation in the system.

An unusual transient feature we observed was three short duration
drops in magnitude observed on night j9 at phases 0.22 (0.04 mag deep
and about 30 minutes in duration), 0.26 (0.02 mag and about 25 minutes
in duration), and 0.31 (0.02 mag and about 45 minutes in
duration). Data from this night are plotted in
Fig.~\ref{drops}. Nothing of the same magnitude is seen on other
nights, or in any of the other stars measured on the same night.  We
have found no indication that they are creations of the reduction
process or were due to poor observing conditions, and the duration is
too short to be related to starspots. One possible explanation is that
star B is a pulsating $\delta$ Scuti variable. The mean photometry for
the system places it within the instability strip (see Sandquist \&
Shetrone 2002), and our deconvolution of the photometric properties of
the three known stars indicates that star B remains in the strip. The
amplitude and short timescale of the features seen is consistent with
$\delta$ Scuti pulsations. Because $\delta$ Scuti stars typically show
multiple modes of pulsation having comparable amplitude, the beating
of the modes can result in brightness fluctuations of several times the
average amplitude for short periods of time. The light from the other
two stars would tend to dilute the apparent fluctuations in the
system's total brightness, perhaps making them unobservable most of
the time. Further monitoring is necessary to prove or disprove this
possibility.

\subsubsection{Starspots and a ``Clean'' Light Curve}

vOVS's model for S1082 accounts for the observed X-ray emission
(Belloni et al. 1993; Belloni, Verbunt, \& Mathieu 1998) as coming
from the interacting binary, and they indicate that it is probably a
variable of the RS CVn type. S1082 is one of the brightest X-ray
sources in the cluster ($L_{x} \sim 7.2 \times 10^{30}$ erg s$^{-1}$),
but has a luminosity consistent with other RS CVns (Singh, Drake, \&
White 1996).  As vOVS also indicate, RS CVn binaries are typically
afflicted with starspots that cause variations in the light curve,
which can potentially explain some of the features of the observed
light curve. Our observations show that the timescale for variations
in the light curve shape is quite short ($\sim$ weeks), which adds to
the evidence in favor of a spot explanation. If correct, then the
spots either have lifetimes (or timescales for change) of less than a
few weeks, or they migrate substantially on week timescales. In the
following we will assume that the variations are indeed due to
spots. There are, however, several unexplained features of the light
curves that may eventually require revisions to the model, as
mentioned throughout \S \ref{analysis}.

Because variations in the light curve are seen at all phases, it is a
considerable task to try to extract robust information from the light
curve. If the light curve shape variations are indeed due to spots,
observations on successive nights can constrain the degree of
synchronism between the rotation of the spotted star and its
orbit. Our best constraints in this regard come from observations of
the secondary eclipse, where we have a series of observations from
successive nights d7/d8, and from nights j1/j3. For d7 and d8, the
overlap runs from approximately 0.4 to 0.5 in phase. On nights j1 and
j3, the overlap is between 0.53 and 0.71. In both cases the
differences between nights generally amount to less than 0.01
mag. Considering that the light curve changed noticeably in the 1.5
months separating the two pairs of nights, the agreement from night to
night appears to indicate a high degree of synchronism on the spotted
star(s).  Because the two stars in the close binary appear to be only
marginally eclipsed (see \S \ref{lcsolve}), this does not directly
show which star is spotted, but it does indicate that the spotted star
is probably synchronous to better than about 3\%. With the rotational
velocity of the star and the binary inclination, this can be used to
constrain the size of the spotted star. If the spots are polar (see
Strassmeier 2000 and Vogt et al. 1999 for some probable examples in RS
CVn systems), this argument will be incorrect. However, polar spots 
would probably be unable to affect the secondary eclipse enough to erase
any hint of contact points.

The only other phase range where we have night-to-night overlap is in
the primary quadrature on nights j7, j8, and j9. Although there is
substantial photometric variation in this portion of the light curve
from month to month (and to some degree during each night), the trends
on these nights agree well. Overall this supports the idea that the
photometric variations are primarily due to just one of the stars (the
rotation velocities of the two stars are too different for them to
both be rotating synchronously; see \S~\ref{rvsec}). The indications
are that the fainter star is the one rotating synchronously (and is
thus the spotted star because of the day-to-day agreement of the light
curves). Unlike most RS CVn systems, the fainter component Ab
contributes a substantial fraction of the light (approximately 25\% of
the light of the close binary), so that activity on the fainter
component could potentially be detectable in the light curve. The
greater degree of asynchronism for component Aa can be used to argue
that its magnetic activity is likely to be smaller than that of component Ab.

The light curve of S1082 bears some resemblance to that of the
short-period ($P = 0.479$ d) RS CVn system XY UMa (see, for example,
Pribulla et al. 2001), where starspots more strongly affect the shape
of the secondary eclipse, and the brightnesses at quadratures
(particularly the secondary quadrature). Although the amplitude of the
variations in S1082 is smaller than in XY UMa, the eclipses are
shallower as well due to smaller orbital inclination and third
light. (Without third light, the primary eclipse would probably be
near 0.27 mag.) As a result, it is very difficult to determine the
shape that the secondary eclipse would have if there were no variability
effects. A number of techniques have been used to try to derive binary
parameters for spotted RS CVn stars (e.g., Budding \& Zeilik 1987),
but significant errors can result if spotting occurs at phases around
the two eclipses, as appears to be the case here.

The method we have chosen to follow to derive a ``clean'' light curve
is a modified version of the method used by Pribulla et al. (2001).
Pribulla et al. assumed that the maximum brightness of a large number
of light curves in every phase bin is the best representation of the
unspotted star, and that the final light curve should be
symmetric. (The existence of substantial hot spots would affect the
validity of the first assumption, as would any variation in the mean
total brightness of the system.)

After our detailed attempts to ensure that all of our data have been
put on the same photometric system, we find that the observations of S
1082 during February and March 2002 were consistently among the
brightest at all phases. As a result, we have used observations from
the three reduced nights as an example of a ``clean'' light curve, as
shown in Fig.~\ref{cleanlc}. The light curve is noticeably asymmetric
about both eclipses, however.

\section{Spectroscopy}

\subsection{McDonald Radial Velocities}\label{mcd}

We have presented results of spectroscopic observations of S1082
previously in \citet{shet00}. After combining some of our highest
resolution spectra to improve the signal-to-noise, we rereduced these
earlier spectra in order to deconvolve the contributions of the three
stars known to contribute to the light.  The spectra were taken using
the McDonald Observatory 2.7~m "2d-coude" \citep{tull95} at resolution
$R=30,000$, and using the McDonald 2.1~m Cassegrain echelle at $R =
43,000$.  These measurements of the radial velocities of the three
known components of the system (which we will refer to as the McDonald
dataset) are listed in Table~\ref{rvs}.

The techniques used to obtain the velocities are similar to those in
\citet{shet00} but the discovery of the previously unnoticed third
component warranted reanalysis.  We used a solar flux template
\citep{hinkle00} to get the velocity of the individual narrow lines.
The velocity error due to the mismatch in temperature between the
template and component B is insignificant due to the small width and
large number of available lines for component B. Cross-correlation was
done with the IRAF task NOAO.RV.FXCOR using high-order continuum
fitting to remove any effect of the broad components.  By shifting the
normalized narrow line components to the heliocentric rest velocity
and combining them we were able to build a very high S/N template to
be used to remove component B from each composite spectrum. (The
removal of each component was accomplished by shifting the median
combined spectrum back to the original observed velocity and then
dividing it into the original unnormalized spectrum.)  Once component
B was removed, the spectrum was again cross-correlated against the
zero-velocity template for component B.  The temperature match between
the Aa and B components is close enough to remove any large systematic
effects.  Once the velocities of component Aa were known, the spectra
were shifted to the heliocentric rest velocity and combined to create
a high S/N template to be used to remove component Aa from each
spectrum.  With both the Aa and B components removed, the velocity of
the Ab component could be determined from cross-correlation against
the zero-velocity B template.  A shifted template of the Ab component
was also created in the same way as the Aa and B templates.  To
determine if the Ab component affected our initial velocities for Aa,
we divided the Ab template into the original composite spectra and
redetermined the velocities for the Aa component.  There was no
significant change in the measured velocities.  A comparison of the
new velocities for Aa and those given in \citet{shet00} shows that all
of the velocities fall within 2 $\sigma$.  The largest deviation (22
\kms) is for HJD 2451620.80, which is the lowest S/N spectrum from the
2.1m telescope.

The rotational velocities ($v \sin i$) were measured using the same
methods described in \citet{shet00}. The values measured for
the close binary were $65 \pm 5$ \kms for component Aa, and $80 \pm 10$
\kms for Ab. The value quoted for Aa is smaller than that quoted in
\citet{shet00} ($90 \pm 20$ \kms) because of the contamination by the
previously unknown component Ab. Our new measurements are consistent
with those of vOVS ($56 \pm 5$ and $83 \pm 5$ \kms,
respectively). 

\subsection{CfA Radial Velocities}\label{cfarv}

As part of a long-term effort to monitor the radial velocities of the
blue stragglers in M67 (Latham \& Milone 1996), we have accumulated
131 spectra spanning 6911 days using the Center for Astrophysics (CfA)
Digital Speedometers (Latham 1992) on the 1.5-m Tillinghast Reflector
and the MMT, both located at the Whipple Observatory on Mt. Hopkins in
Arizona. The instruments on the two telescopes were almost identical
and used intensified photon-counting Reticon detectors to record 45 \AA~
of spectrum in a single echelle order centered at 5187 \AA~ with a
resolution of 8.5 \kms.  The typical signal-to-noise ratio for these
spectra is about 20 per resolution element, but a few of the earliest
exposures have signal-to-noise ratios below 10.

Radial velocities were extracted from the observed spectra using the
one-dimensional correlation package {\bf r2rvsao} (Kurtz \& Mink 1998)
running inside the IRAF environment.  For the templates we
used our new library of synthetic spectra (Morse \& Kurucz, in
preparation) calculated using the latest Kurucz model atmospheres.

To select the optimum effective temperature and rotational velocity
for our synthetic template we ran correlations for a grid of
templates, assuming solar metallicity and log $g = 4.0$ for the
surface gravity.  This analysis gave the highest average peak
correlation value for the template with $T_{\rm eff} = 6750$ K and $v
\sin i = 12$ \kms (a quadratic interpolation incorporating the results
from the adjacent templates gave $T_{\rm eff} = 6866$ K and $v \sin i
= 13.6$ \kms).  The individual radial velocities derived using this
template are reported in Table~\ref{cfarv.tab}, where we list the
telescope used (T stands for Tillinghast Reflector and M for MMT)
heliocentric Julian Day, and the radial velocity, estimated error from
{\bf r2rvsao}, and $o-c$ velocity residuals from the orbital fit
reported below, all in \kms.  The estimated error includes a floor
error contribution of 0.25 \kms, combined with the internal error
estimate from {\bf r2rvsao} in quadrature.  This 
floor error is needed to compensate for the various systematic errors
that are not taken into account by {\bf r2rvsao}, such as small
changes from night to night in the zero point of the absolute velocity
system that are not fully corrected for (e.g. Latham et al.\ 2002).
The velocities reported in Table~\ref{cfarv.tab} are on the native CfA
system using nzpass=2; 0.139 \kms should be added to these velocities
to put them on an absolute system defined by extensive CfA
observations of minor planets.

In Figure~\ref{cfarv.fig} we plot the history of the CfA velocities
for component B (upper panel) and the corresponding power spectrum
(lower panel). Although the RMS deviations of the velocities from the
mean value are only 1.70 \kms, this is significantly larger than the
average internal error estimate of 0.95 \kms.  In particular
P($\chi^2$), the probability that a star with constant velocity could
give by accident a value of $\chi^2$ equal to or larger than the
observed value of $\chi^2=464$, is very small, less than
$10^{-6}$. Indeed, the power spectrum shows a strong peak at a
frequency corresponding to a period of about 1200 days. The velocity
variations with this period are not easy to see in the plot of the
velocity history except for the most recent cycle, where we made an
effort to get strong exposures with good time coverage.

The CfA velocities for S1082 lead to an orbital solution with period
$P = 1188.5 \pm 6.8$ days.  The orbital amplitude $K = 2.28 \pm 0.26$
\kms is small, but the uncertainty in $K$ is almost ten times smaller.
In our experience orbital solutions with this large a ratio for the
uncertainty in $K$ and spanning several cycles are robust.  We looked
for orbital solutions at other periods in the range 0.9 to 10,000 days
without success.  The orbital parameters for our solution are listed
in detail in Table~\ref{cfaorb.tab} and the velocity curve is plotted in
Figure~\ref{cfaorb.fig} along with the observed velocities. While 
signatures of the components of the close binary are present, additional
analysis is underway to extract the parameters of that orbit.

The systemic velocity that we find, $\gamma = 33.76 \pm 0.12$ \kms, is
nicely consistent with the cluster mean velocity of 33.5 \kms and
cluster dispersion of 0.5 \kms reported by Mathieu \& Latham (1986).
Although that result was based on observations made with the
CfA Digital Speedometers, the procedures used for deriving the
individual velocities and setting the absolute velocity scale have
evolved considerably over the intervening years.  A definitive
reanalysis of the cluster membership, mean velocity, and velocity
dispersion is underway.

\subsection{Surface Temperatures}\label{tsecm}

Our high resolution spectroscopy for this system can provide valuable
guidance on the surface temperatures of the three components,
independent of the photometry. Because of the sizable distortions of
the light curve of the system, it is worthwhile to consider the
possibility that starspots have affected our temperature
measurements. Because some of our spectra were taken during the same
period as our photometry we can gauge of the nature of the spot
activity at the phase of observation.  The temperature for the narrow
component B was determined from the McDonald spectra following
\citet{shet00}. Briefly, the temperature was determined by measuring
the abundance from many different iron lines of different strengths,
excitation potentials, and ionizations and then minimizing the
abundance differences measured from different lines.  Using this
technique the best temperature $T = 6950 \pm 100$ K, surface gravity
$\log g = 4.3 \pm 0.15$, and microturbulent velocity $v_{t} = 2.1 \pm
0.1$ \kms were found to yield a number abundance of $\log N($Fe$) =
7.46 \pm 0.03$ ($\log N($H$) = 12.0$). Forcing that abundance to equal
the cluster mean ([Fe/H] $= -0.04$) would yield a flux ratio $F_{B} /
F_{tot} = 0.63 \pm 0.02$.  This technique works well for narrow-line
stars for which many cleanly measurable Fe lines can be used.  For
rapidly rotating stars, the lines are too blended to measure the
equivalent widths of individual lines, so that a different technique
had to be employed.

We looked for spectral regions within a single echelle order that
contained lines with different sensitivities to temperature. A
comparison of the depths of these blended lines could be used to set a
temperature scale.  This is similar to the techniques of
Gray \& Johanson (1991; refined in Gray 1994), who used individual line
depths to very accurately measure temperatures.  Instead of using the
zero-velocity templates for Aa and Ab created in the radial velocity
analysis, we created another template using six spectra from December
6, 7, and 13, 2000 (which were taken within days of one of our
photometric runs). The procedure used to isolate each component was the same as the procedure
described in \S \ref{mcd}.   No effort was made to weight the data
by phase. As seen in Fig.~\ref{allphot}, our December 2000
observations agreed well with our observations of February and March
2002, when the system was at its brightest. Phases near the primary
quadrature and leading into secondary eclipse appear to be an
exception however: there is a noticeable asymmetry in the eclipse, and
the early portions appear to be low by about 0.02 mag. Four of our six
spectra were taken in the secondary quadrature, which appears to have
been unaffected. A fifth observation appears to have been taken at a
phase (0.434) when the photometric effects are relatively small ($\sim
0.01$ mag). The sixth observation appears to have been taken at a
phase ($\phi = 0.294$) when the light curve shows a larger deviation.

The spectral regions that were selected were 4555 -- 4595 \AA ~and
5250 -- 5330 \AA ~(within the $B$ and $V$ passbands, respectively) for
relatively weak contributions from the narrow line component B
and relatively strong contributions from the broad line components.
Lines that are relatively insensitive to temperature were used to
measure relative fluxes of the three components.  To set the
temperature scale we used two techniques. The first was based upon the
colors and spectra of other observed blue stragglers in the sample of
\citet{shet00}. We converted the colors to temperatures using equation
3 in Soderblom et al. (1993).  We then broadened spectra to the width of
both components Aa and Ab using an elliptical rotation kernel.  This
allowed us to measure line depths in blue stragglers with known colors
(temperatures) and compare them with the line depths in Aa and Ab. The
drawback of this method was the small number of comparison stars.  

Our second method used synthetic spectra with temperatures between
5500 K and 8500 K for the chosen regions to determine line ratios at
each temperature.  This provides excellent temperature coverage but is
limited by model inaccuracies such as line oscillator strengths and
model parameters. An example of the second method applied to one
spectral feature is given in Fig.~\ref{syn}. The strong feature at
4563.9 \AA ~is dominated by \ion{Ti}{2} and the strong feature at
4565.2 \AA ~is a blend of two strong \ion{Fe}{1} lines and one
\ion{Cr}{1} line.  In the synthesis using $T_{eff} = 8500$ K, the
4563.9 \AA ~line is much stronger than the 4565.2 \AA ~feature.  In
the 5500 K synthesis the 4565.2 \AA ~feature becomes stronger than the
4563.9 \AA ~feature.  The relative strengths of these two features
gives us a temperature indicator.  In the spectrum of Ab the 4565.2
and 4563.9 \AA ~features have roughly comparable strengths, implying a
temperature of about 6000 K.

By combining the results from these two methods we
hope to reduce the errors and produce an accurate estimate of the
temperature. We also used the absolute line depths of
temperature-insensitive blends to measure the relative flux ratios.
By using the fact that the total flux contribution should add up to
1.0 we further constrain the flux ratios and derived temperatures.  In
Table~\ref{ts} we present the results of our analysis for each region.
The agreement between the different methods and regions is quite good
for the Ab system, while the agreement among the Aa estimates is not
as good and thus has a large associated error.

One additional constraint on the temperatures of the components of S
1082 comes from photometry. Because the main goal of our photometry
was light curve coverage, we focused on observations in $V$ band.
vOVS use the photometry of Montgomery et al. (1993),
although those data may not have had enough coverage to give a good
indication of the average magnitude of the system. The Montgomery et
al. paper gives $V = 11.251$ and $(\bv) = 0.415$. Because the best
model of vOVS gives $\bv = 0.48$, there is still room for improving
the system parameters. Using the reddening and temperature-color
conversions of Soderblom et al. (1993) we find that our best estimate
"Combined" temperatures produce a system that is too blue, $(\bv) =
0.362$.  However, taking into consideration the errors in flux and
temperature, a good fit can be obtained by using the coolest
temperature within the errors provided.  For $T_{Aa} =$ 7325 K with
$F_{Aa} / F_{tot} = 0.30$ and $T_B =$ 6850 K with $F_B / F_{tot} =
0.63$, the combined system color is $\bv = 0.409$.  (Because of the
low flux contribution from the Ab component the system color does not
provide much constraint on its temperature.)  The fluxes and
temperatures based upon this analysis are listed in the ``Constrained''
row of Table~\ref{ts}. We also note that this determination of $T_B$
is consistent with the value derived independently from the CfA data,
as described in \S \ref{cfarv}.

\section{Discussion}

\subsection{Surface Temperatures and Relative Fluxes}\label{tsec}

Our temperatures for the stars in the close binary are substantially
higher than the temperatures quoted by vOVS, which were derived from
photometry and low-resolution spectra. However, because the relative
depths of the eclipses in the light curve constrain $T_{Aa} / T_{Ab}$,
our value for the temperature ratio is consistent with that of
vOVS. The small difference in the temperature ratio could easily
result from systematic error in our spectroscopic temperatures, spots
affecting the photometric solution of vOVS, or incomplete separation
of the different components in the low resolution spectra of
vOVS. vOVS also indicate that their photometric solution is somewhat
uncertain when it comes to the relative fluxes of components Aa and
B. The observational constraint from our spectra favors putting 61\%
of the flux in component B (as opposed to 42\% for vOVS).

The combination of effective temperature and relative fluxes allows us
to constrain the color-magnitude diagram positions of the component
stars, given the system's total photometry. As seen in Figure~\ref{cmd}, the
components of the close binary are both found near a zero-age main sequence:
component Ab is on the fiducial line of the cluster below the turnoff,
while component Aa falls near the zero-age main sequence (ZAMS),
clearly in the blue-straggler region. Component B is slightly redder
than the total color of S1082, and very clearly to the red of the ZAMS.

\subsection{Light-Curve Solution}\label{lcsolve}

The light-curve models of vOVS were computed with the ELC code of
\citet{oro00}, which uses recent stellar-atmosphere models and a
sophisticated algorithm to search for the global $\chi^{2}$ minimum
fit to their photometry and radial velocities.  The vOVS photometric and
radial-velocity datasets had significant gaps in coverage,
however. One effect of this was that they were forced to assume that
the stars in the close binary were free of spots. Because we have much
more extensive photometry in the $V$ band and additional constraints from
spectroscopy, we have modeled the system independently.

In finding a solution to the light curve, we have attempted to rely on
our observational constraints as much as possible.  The reason for
this is that even with light curves in several wavelength bands and
radial-velocity curves, vOVS found that there were a ``relatively
large number of solutions with similar $\chi^{2}$ values''. 

In using the ephemeris for the close binary derived by vOVS, we found
it necessary to shift the data by 0.005 in phase to get an adequate
fit for the eclipses in the 2002 data. This shift is well within the
observed minus computed (O-C) values for time of primary minimum found
by vOVS, so we feel this is justified. Our spectra not only give us
the first direct constraints on the absolute temperatures of the three
components of the system, but they also give us a consistency check on
the relative eclipse depths in the light curve (which constrains
$T_{Aa} / T_{Ab}$). Our relative flux measurements also directly
constrain the amount of ``third light'' in the solution, which reduces
the systematic uncertainty in the inclination and Roche lobe filling
factors (which affect the depths of both eclipses relatively equally).
We take the mass ratio $q$ to have the value derived by vOVS from
their radial velocities. (Our data do not give any indications of a
need to change this value.) Comparing our best model using the new
constraints to the best model of vOVS will provide an indication of
what characteristics of the system are most robustly measured at this
point.

In our fits we used the program NIGHTFALL\footnote{See
http://www.lsw.uni-heidelberg.de/$\sim$rwichman/Nightfall.html for the
program and a user manual (Wichmann 1998)}. We verified the code by
reproducing the solution of vOVS on their data before proceeding. The
NIGHTFALL code allows for effects such as detailed reflection,
asynchronism of the stars (and the corresponding modification to the
star shapes), model atmospheres (Hauschildt et al. 1999), and third
light. The eccentricity of the binary was always assumed to be
zero. Table~\ref{lccomp} list the parameters of our best fit to the
``clean'' light curve of the February/March 2002 data.  Of particular
importance is the inclination $i$, which we find to be only a few
degrees higher than the value given by vOVS. Large systematic error is
not to be expected, given that the inclination is constrained by
eclipse depth and duration.

The lobe filling factors for the stars in the close binary are only
directly constrained by the light curve, but this does restrict the
stars to being well within their respective Roche lobes. Despite the
differences between our input parameters and those of vOVS, we find
that the filling factor of component Aa agrees very well with theirs
(the inclusion of asynchronous rotation has an effect of a few
percent).  The filling factor we derive for component Ab is
larger than that derived by vOVS, indicating that the
star is closer to filling its tidal lobe. The difference in the two
studies appears to come from our constraint on the ``third light'' of
the system: the larger contribution from component B implies shallower
eclipses, which in turn requires increased star size to bring the
light curve back to what is observed.

An important feature of the light curve is the width of the secondary
eclipse, which appears larger than the width of the primary
eclipse. We have modeled this feature with a cool spot on the side of
the secondary star facing the primary (a longitude of $187 \degr$,
facing us shortly after the secondary eclipse). The eclipses observed
on nights f1.2 and f6.2 are very nearly symmetric, as is the eclipse
observed in January 2001. Observations by us in December 2000, March
2001, and by vOVS in 1999 and 2000 show larger degrees of asymmetry,
however. These variations in the shape of the secondary eclipse lends
credence to the idea that starspots are important, and disputes
the idea that ellipsoidal variations alone are the cause (as is the
case in the vOVS light curve solution). In any case, our fits of the
2002 data imply a spot of fairly large size ($\sim 30 \degr$) and
temperature decrement ($T_{spot} / T_{photo} \approx 0.8$), although
there are other combinations that also fit reasonably well.

If this were the only spot on the system, then presumably phases near
primary eclipse should be fit well. There is, however, a noticeable
asymmetry in the light curve on either side of the eclipse, with
phases $0.08 < \phi < 0.25$ generally being slightly fainter than the
$0.75 < \phi < 0.92$. As a result we believe it is necessary to
hypothesize additional spots on the fainter star Ab. It does not seem to be
possible to explain the light-curve shape with a single spot. We
placed the additional spot (size $\sim 20\degr$ and $T_{spot} /
T_{photo} \approx 0.9$) at longitude $40 \degr$ (approximately midway
between primary eclipse and the following quadrature).

We have not done a detailed analysis of the random error in our light
curve model because there is clear evidence that systematic errors are
of more importance to the uncertainty. Although our $V$ band curve is the
most complete light curve that has been presented for this system to
date, it is still unclear exactly what is the ``undisturbed'' light
curve. In addition, there are still unexplained features (see \S
\ref{analysis}) that could have a bearing on the final fitted
parameters.

\subsection{The Orbital Solution}

Previous studies (Goranskij et al. 1992; Shetrone \& Sandquist 2000)
failed to extract an orbital period from the radial-velocity data of star B
alone. As was pointed out by vOVS, the radial-velocity variations are
inconsistent with star B being in the close binary.

\subsubsection{The Possibility of a Triple System}

Proper motions have established that the star that is providing most
of the light is a member (94\%, Sanders 1977; 99\%, Girard et
al. 1989; 92\%, Zhao et al. 1993). The inference is that S1082 is
either a physical triple system or a chance superposition of
relatively bright cluster members.  Given that S1082 was identified as
a blue straggler (and it is indeed bluer than the cluster turnoff), it
is necessary to have a blue straggler as one of the three stars that
is observable. Indeed the models of S1082 presented by vOVS and by us
below require that both the narrow line component B and the broad line
component Aa are blue stragglers in their own rights.

We are unable to detect a reflex motion in the mean radial velocity of
the close binary system because of the weakness of the broad-line
components, although it is fairly clear that the close binary is a
member of M67 (see \S \ref{closeb}). If the close binary is
dynamically associated with star B, then the reflex motion is also
likely to be less than the orbital amplitude $K$ of star B because of
the large combined mass of the binary. As a result, it is not possible
at this time to rule out the possibility that the narrow component is
part of a binary that is superimposed on the close binary by
chance. (A partial overlap is more likely than an unresolved blend,
but there is no evidence of two peaks in the photometric data, and
there are currently no {\it Hubble Space Telescope} observations of
this system that can be used to more tightly constrain whether the
system is a blend of non-interacting systems.) The small $K$
semi-amplitude for component B (in the nomenclature of vOVS) indicates
that the inclination of the system must be very low if it is indeed
part of a hierarchical triple. If we take the vOVS model of S1082 to
have the correct masses, the probability of such a small inclination
occurring by chance is approximately 0.3\%.

We can estimate the probability of superposition of two blue
stragglers in M67. We conducted Monte Carlo experiments, randomly
selecting the positions of stragglers relative to the cluster center
in polar coordinates. The angular coordinate was selected randomly,
and the radial coordinate was selected using the observed cumulative
radial distribution of the 29 identifiable blue stragglers in M67
(Sandquist \& Shetrone 2002) as the underlying
radial distance probability distribution. 29 stragglers were selected
in each trial, and superpositions with angular separations of less
than one arcsecond were noted. By this method, we found a probability
of approximately 0.4\% that two stragglers would be
superimposed. Thus, S1082 presents us with two possible explanations
that have fairly low (but nearly equal) probabilities of occurrence.

We have two other pieces of circumstantial evidence that bear on this
question. First, the eccentricity of the orbit of star B is measured
from the orbital solution to be $e = 0.568 \pm 0.076$. As the
distribution of eccentricities resulting from binary interactions is
expected to be thermal ($dN/de \propto e$), such a large eccentricity
for the outer orbit is most consistent with a dynamical formation
scenario. However, this does not rule out the possibility that a two
star binary containing star B and a low-mass companion was formed
independently of the close binary A by dynamical processes, and is
responsible for the radial-velocity variations seen today for star B.

We can also place some limits on variations in the timing of the
eclipses that could result from reflex motions of the close binary and
light travel time (vOVS). To test this, we used our best theoretical
light curve to simultaneously fit the mean light level in the $V$ band
and the O-C time of primary eclipse compared to the ephemeris of
vOVS. Given the small size of light curve variability on month
timescales and the long period of the outer orbit, we combined subsets
of our photometric data into partial light curves for December 2000,
January 2001, February/March 2001, and February/March 2002.  In the
resulting $\chi^{2}$ maps, we found that while the mean light level
was well constrained, the (O-C) times were not. Although our
observations cover the outer orbit from phase 0.05 to 0.45 (which
should have nearly spanned the maximum variation in eclipse timing),
we are only able to put an upper limit of 0.01 d on the
variations. The presence of spots was a clear systematic effect on the
measured (O-C) values, particularly because the light curve of the
secondary eclipse at different epochs was often asymmetric around
phase 0.5 and sometimes the minimum was shifted away from phase
0.5. If a more concerted effort to look for light travel time
variations is to be made, the effort should be concentrated on
observations of the primary eclipses since
this feature seems less sensitive to spotting. In any case, the
current dataset is not sufficient to detect reflex motion of the close
binary. As a result, there is still not definitive evidence indicating
a dynamical link between star B and the close binary.

\subsubsection{The Close Binary}\label{closeb}

The radial-velocity measurements from this paper and from vOVS are
shown in Fig.~\ref{rvmeas}.  Many of the spectroscopic observations in
the McDonald dataset were taken at binary phases between 0.4 and 0.9,
which is coincidentally a range of phases where vOVS did not have good
coverage. Unfortunately, it becomes more difficult to adequately
deconvolve the signatures of the two broad-line components when they
are near eclipse, so that these measurements do not improve on the
orbital solution of vOVS. However, they do help to establish that the
binary (and most clearly, component Aa) is probably a member of the
cluster.

As emphasized by vOVS, a better determination of the eccentricity of
the close binary is important for understanding the system.  Additional
observations of higher quality taken at phases between 0.6 and 0.9 would
provide the best constraints. The timescale for damping the
eccentricity of a binary orbit with a period near a day is expected to
be so short that a detection would most likely indicate pumping of the
binary's eccentricity by component B (e.g. Mardling \& Aarseth 2001),
providing evidence of a dynamical association between the stars.

\subsubsection{Rotational Velocities}\label{rvsec}

With the information on synchronism from starspots and
constraints on the inclination from the light curve, we can compute
the radius of component Ab. We assume that the inclination of the
spotted star's spin axis is the same as the inclination of the orbit,
although for most asynchronous RS CVn systems this is not the case
(Glebocki \& Stawikowski 1997). Because S1082 appears to be a system
that is {\it mostly} synchronized, the assumption about the
inclination of the star's spin axis is probably reasonable. Our values
give an equatorial radius of $R_{Ab} / \rsun = 1.82 \pm 0.23$. This is
lower than our values derived from the light curve in \S \ref{mr},
which probably indicates a small degree of asynchronism or
misalignment of the spin axis with the orbital plane.

The model fits of vOVS are consistent with an orbital eccentricity of
zero, but the stars have spin factors $\Omega = \omega_{rot} /
\omega_{orb}$ less than one. Our measurements also indicate both stars
are rotating asynchronously (see Table \ref{phys}), although the
differences in our rotation velocities make it appear that the two
stars are asynchronous to nearly the same degree. (The repeatability of
the light curve from day to day can be used to constrain the degree of
asynchronism, although the spotted star must be more definitely identified
first.) Because of the asynchronism, we infer that the close binary is
relatively young from a dynamical standpoint. In M67, the maximum
period for a tidally-synchronized binary is 10.3 d
\citep{mathieu90}. Most RS CVn systems in the field with periods less
than this are synchronized (Glebocki \& Stawikowski 1995), with only
one asynchronous RS CVn star known having a period shorter than S1082
(IL Com; $P = 0.96$ d).

As vOVS point out, a third star can induce eccentricity in the orbit
of a close binary, or desynchronize the rotation of one or both stars
in the binary. However, because of the much larger moment of inertia
for the binary compared to the individual stars, a measurable
eccentricity is a much more likely result of interaction with a third
star than is asynchronism. Even if there is a small but non-zero
eccentricity, a short-period binary like S1082 should be
pseudo-synchronized with the orbit.

\subsubsection{Component Masses and Radii}\label{mr}

Our light-curve models indicate that the inclination of the close
binary is likely to be only a few degrees larger than the value
presented by vOVS. Our final derived values for the physical
parameters of the stars are shown in Table \ref{phys}.  The small
reduction in the mass of Aa means that it is more likely that the star
formed from the coalescence of just two turnoff-mass stars, rather
than three as required if the value derived by vOVS is correct
(although their quoted error bars are consistent with twice the
turnoff mass). The reduction in mass, however, is not nearly enough to
explain the strong disagreement between the inferred masses of the
stars and stellar-evolution predictions for the positions of these
stars in the color-magnitude diagram (CMD). Component Ab would need to
be approximately 2 magnitudes fainter than a single star of the
computed mass in M67. Component Aa would need to be significantly
fainter still.

vOVS explored and rejected the possibility that the close binary is
not a member of the cluster. Indeed, this would make it difficult to
explain the fact that both components appear to fall near an extension
of the main sequence for the cluster. In addition, a larger distance
modulus would require a substantial increase in the calculated X-ray
luminosity ($L_{X}$), meaning that the binary could probably not be of
the RS CVn-type, and almost certainly not an Algol because they tend
to have smaller $L_{X}$ for a given period (Singh et al. 1996). More
importantly, the systematic velocities of the stars appear consistent
with that of the cluster to within the measurement errors.

Although the values for the two radii are consistent with each other
within the errors, we find, as did vOVS, that the evidence indicates
that the radius of component Aa is smaller than the radius of the
less-massive component Ab. For coeval stars, this should not be
possible. It is possible to construct reasonable light-curve models if
component Aa is forced to be the larger of the two stars, and filling
more of its Roche lobe than component Ab. Although this makes the two
stars appear more nearly coeval, this would cause the width of the
primary eclipse to greatly exceed what is observed. In addition, both
stars would be noticeably out of synchronization with the orbit, in
contradiction to the evidence indicating that the shape of the light
curve repeats from night-to-night over timescales shorter than a few
weeks.

\subsubsection{Consistency}

Although we heavily used spectroscopic information to constrain our
light-curve fitting, we can check to see if the light-curve solution
is consistent with {\it all} of the spectroscopic data. The light
curve most tightly constrains the sum of the stellar radii, while the
stellar fluxes derived from spectroscopy constrain the individual
radii, assuming the temperatures are well determined. Our best fit
light-curve model predicts a higher value for the flux ratio $F_{Ab} /
F_{Aa} = 0.4$ than derived from the spectroscopy ($0.30 \pm 0.07$). This
can be resolved by reducing the radius of star Ab and increasing the
radius of star Aa. This goes in the direction of bringing star Ab
closer to synchronous rotation, as required by the day-to-day
repeatability of the light curve.

Slight adjustments do need to be made to other light-curve parameters
to improve the fit to the light curve if these changes to the radii
are made.  The indications are that the inclination needs to be
increased slightly ($i \sim 68\fdg5$) and the temperature dimming
factor of the spot on the secondary ($T_{spot} / T_{photo} \sim
0.7) $ needs to be reduced. While the increase in inclination reduces
the component masses and makes a binary-binary collision scenario more
believable, the size of the temperature decrement due to the spot
begins to become physically unrealistic ($\sim 1800$ K). Further
modeling of S1082 should be done to attempt to eliminate these
inconsistencies.

\section{Conclusions}\label{conc}

Our study of the light curve for the close binary in S1082 supports
the hypothesis put forth by vOVS that the close binary is an RS CVn
system, and radial velocities presented here establish that the
brighter third component is part of a multiple-star system itself.  To
this point though, there is no direct evidence proving a dynamical
link between the close binary and the component B. Although components
A and B have systematic velocities that indicate they are both cluster
members, the small amplitude of the radial velocity curve for B would
require a low-probability, nearly face-on orientation of its
orbit. Our new light-curve solution results in small changes to the
radii of the stars in the close binary as compared to the solution of
vOVS, but approximately 10\% reductions of the derived masses. The new
determinations of radii and masses agree with those of vOVS to within
the errors. As a result, we have been unable to reconcile the implied
masses and radii of the stars in the close binary with their
color-magnitude positions. For component Aa, its radius seems to be
smaller than its lower mass companion Ab.

If they were found separately in the cluster, both the close binary
and component B would be identified as blue stragglers. This fact
appears to be independent of the current uncertainties in the
brightness characteristics of the stars.  So what does this system say
about blue stragglers? The orbital solution of the close binary
implies that it contains two stars of greater than turnoff mass, which
in itself rules out a mass transfer explanation and strongly supports
collisional formation.

Clearly S1082 would be an important system if a dynamical link could
be established between components A and B since the formation of the
system would require at least 5 stars (vOVS) and more likely 6,
more or less requiring multiple binary-binary interactions. Is this
consistent with what we know about the system? First, it has been
established by several authors that the rotation rate for component B
is less than 20 km s$^{-1}$. If the straggler formed by a collision
between two of the stars in the input binaries, the straggler would
most likely have received substantial angular momentum, although it is
currently unclear how rapidly the angular momentum would have been
lost (Sills et al. 2001). Any attempts at age-dating component B using
rotation are complicated by tidal effects in the close binary that
have apparently erased signs of the original angular
momentum. However, it is significant that both components appear to be
rotating {\it slower} than synchronous rotation. This may indicate
that the spin-down timescale for both stars was much shorter than the
synchronization timescale, or that the close binary was formed after
the individual stragglers (which seems unlikely given their relatively
short lifetimes).

Tidal effects in the close binary probably constrain the dynamical age
of the close binary to be small, since the binary is effectively
circularized (eccentricity consistent with zero) but at least one of
the stars is far from rotating synchronously.  Because of the
magnitude of the tidal effects in the binary, it is likely that
component B formed first, regardless of whether it is or is not
dynamically associated with the binary. In our model for the close
binary, even the minimum mass of component Aa ($(M_{Aa} /
\msun) \sin^{3} i = 2.01 \pm0.38$; vOVS) puts it beyond the point at
which it should have a surface convection zone. Models that put
component Ab on the main sequence (our model) or on the subgiant
branch (vOVS) might lead us to expect a surface convection zone is
present. The calculated masses of the stars in the close binary are
inconsistent with their CMD positions though, which makes details of
the interaction tricky to interpret. Star Ab in particular has
characteristics that put it in a regime where tidal dissipation theory
is uncertain, and where the stellar structure becomes critical. The
minimum mass for star Ab ($(M_{Ab} / \msun) \sin^{3} i = 1.26 \pm
0.27$; vOVS) would imply a small surface convection zone while the
modeled mass ($M_{Ab}$) would imply a radiative envelope. The ratio of
the synchronization timescales for the two stars does alleviate some
of the theoretical uncertainty. Because of the similarity in the
stellar radii, star Ab would be expected to take substantially less
time to synchronize than star Aa unless Ab has a radiative envelope
(e.g. Tassoul \& Tassoul 1992, Claret \& Cunha 1995). If the
rotational velocities of the two stars can be measured more
accurately, this may allow us to probe the energy-transport mechanism
dominating in the envelope of star Ab.

The best lower limit on the age of the close binary probably comes
from the thermal timescale for relaxation ($\la 10^{7}$ yr) after a
collision \citep{sills01}.  An upper limit for the evolutionary
timescale of the brighter component of the binary can be derived from
the lower limit on the mass of star Aa. Blue stragglers that form via
stellar mergers tend to wind up with the core of the more evolved star
at the center of the remnant. If star Aa was formed by the merger of
two equal mass stars of about $1 \msun$ (consistent with the minimum
mass), the remnant would have the maximal core hydrogen fusion
lifetime. An additional factor prolonging the lifetime of a blue
straggler of this mass would be the creation of a convective core that
would help mix hydrogen into the core. In fact, if our model mass
($M_{Aa}$) is correct a convective core makes it possible to form this
star from just two near-turnoff mass stars. It is even feasible for the
input stars to have had masses slightly higher than turnoff mass
because the convective core would significantly extend the lifetime of
the straggler formed. The resulting straggler should then have had a core
hydrogen-burning lifetime of between about 0.3 and 0.6 Gyr (Schaller et
al. 1992).

%This limit would not hold if the cores of more than two stars were
%incorporated into the more massive component of the binary. If there
%are not substantial systematic errors in the mass ratio and
%inclination of the close binary, then three stars appear to be needed
%to form component Aa. In such a case, the zero-age main sequence
%lifetime is the safest upper limit for its age. Even though the
%highest mass input star would tend to become the core of the
%straggler, convection would tend to limit the initial core helium
%abundance. The age upper limit would then be under 0.6 Gyr. The
%apparent positions of both components near the ZAMS probably indicate
%that this is an overestimate of the age, consistent with dynamical
%indications.

\acknowledgments

We would like to thank M. van den Berg and K. Stassun for providing
data from their study for use in the comparisons presented above, and
M. van den Berg for refereeing the paper.  E.L.S. would also like to
thank R. Taam, P. Etzel, J. Wood, and P. Heckert for valuable comments
during the course of this work.

\begin{figure}
%\plotone{f1.eps}
\caption{The $V$ light curve of S1082 phased to the ephemeris of van
den Berg et al. (2001). Night identifications are listed in
Table~\ref{obs}. Symbols are grouped by month of observation: filled
symbols (December 2000), skeletal symbols (January 2001), open symbols
(February and March 2001), and stellated symbols (February and March
2002). Typical errors were 0.003 to 0.007 mag. \label{allphot}}
\end{figure}

\begin{figure}
%\plotone{f2.eps}
\caption{$V$-band data for S1082 separated by month. Top panel: data
from van den Berg et al. 2001. In all other panels, the points have
the same meaning as in Fig.~\ref{allphot}. \label{monthphot}}
\end{figure}

\begin{figure}
%\plotone{f3.eps}
\caption{$V$-band data for S1082 on night j9 (Jan. 31/Feb. 1,
2001). \label{drops}}
\end{figure}

\begin{figure}
%\plotone{f4.eps}
\caption{The fit to $V$ band data for S1082 from February/March 2002.
The points have the same meanings as in Fig.~\ref{allphot}. \label{cleanlc}}
\end{figure}

\begin{figure} 
%\plotone{f6.eps} 
\caption{The CfA radial-velocity history (upper panel) and
corresponding power spectrum (lower panel) for the narrow-lined star
in the outer orbit. \label{cfarv.fig}}
\end{figure}

\begin{figure} 
%\plotone{f7.eps} 
\caption{The velocity curve for the orbital solution together with the
observed velocities for the narrow-lined star in the outer orbit
\label{cfaorb.fig}}
\end{figure}

\begin{figure}
\caption{An example of one (of several) of the temperature-sensitive
regions used in our spectral temperature analysis.  The solid line
represents the composite spectrum of Ab, the dotted, short-dashed,
long-dashed and dash-dotted lines represent synthetic spectra with
effective temperatures of 5500, 6500, 7500 and 8500 K.  We have offset
the composite spectrum of Ab to cleanly separate it from the
syntheses, but no offset has been made between the syntheses. The
stellar spectrum has been stretched to account for the diluting effects
of the other two components in the spectrum.
\label{syn}}
\end{figure}

\begin{figure}
%\plotone{f5.eps}
\caption{Color-magnitude diagram of M67 (Fan et al. 1996) with open
points showing our computed positions for the components of S1082
({\it squares}: components of the close binary; {\it triangle:}
component B; {\it filled square:} the model values for the blend of
the three stars.\label{cmd}}
\end{figure}

\begin{figure}
%\plotone{f8.eps}
\caption{Radial-velocity measurements from this study ({\it filled
symbols}) and from vOVS ({\it open symbols}). All three measured
components are shown: component Aa ({\it triangles}), Ab ({\it
squares}), and B ({\it circles}). The fit to the close-binary
components by vOVS ($q = 0.63$, $e = 0$) is shown with the solid
lines.\label{rvmeas}}
\end{figure}

\begin{deluxetable}{ccccccccc}
\hspace*{-0.2in}
\tablewidth{0pc}
\tablecaption{Observing Log for $V$ Photometry at Mt. Laguna}
\tablehead{\colhead{Local Date} & \colhead{} & 
\colhead{mJD Start\tablenotemark{a}} & \colhead{$N_{obs}$}}
\startdata
Dec. 5/6, 2000 & (d1) & 1884.878 & 60\\
Dec. 7/8, 2000 & (d3) & 1886.806 & 111\\
Dec. 11/12, 2000 & (d7) & 1890.817 & 57\\
Dec. 12/13, 2000 & (d8) & 1891.820 & 116\\
Jan. 23/24, 2001 & (j1) & 1933.675 & 206\\
Jan. 25/26, 2001 & (j3) & 1935.673 & 176\\
Jan. 29/30, 2001 & (j7) & 1939.676 & 44\\
Jan. 30/31, 2001 & (j8) & 1940.657 & 130\\
Jan. 31/Feb. 1, 2001 & (j9) & 1941.663 & 161\\
Feb. 17/18, 2001 & (f1) & 1958.655 & 63\\
Feb. 18/19, 2001 & (f2) & 1959.790 & 75\\
Mar. 1/2, 2001 & (m1) & 1970.619 & 205\\
Mar. 3/4, 2001 & (m3) & 1972.631 & 205\\
Mar. 5/6, 2001 & (m5) & 1974.721 &  32\\
Feb. 5/6, 2002 & (f1.2) & 2311.649 & 119\\
Feb. 10/11, 2002 & (f6.2) & 2316.638 & 167\\
Mar. 18/19, 2002 & (m1.2) & 2352.625 & 131\\
\enddata
\label{obs}
\tablenotetext{a}{mJD = HJD - 2450000}
\end{deluxetable}

\begin{deluxetable}{cccrrcc}
\hspace*{-0.2in} 
\tablewidth{0pc}
\tablecaption{Radial-Velocity Measurements from McDonald Observatory}
\tablehead{\colhead{HJD} & \colhead{Phase} & \colhead{$v_{r,B}$} &
\colhead{$v_{r,Aa}$} & \colhead{$v_{r,Ab}$} & \colhead{S/N} & 
\colhead{Source}}
\startdata
2450828.88 & 0.889 & $32.8 \pm 0.8$ & $51 \pm 3$   & $-84 \pm 5$   & 74 & 2.7 m\\
2450855.62 & 0.931 & $34.9 \pm 1.2$ & $31 \pm 10$  & $-73 \pm 20$  & 11 & 2.7 m\\
2451619.67 & 0.470 & $34.0 \pm 0.3$ & $31 \pm 5$   & $-30 \pm 10$  &146 & 2.1 m\\
2451619.74 & 0.535 & $34.0 \pm 0.4$ & $22 \pm 5$   & $-50 \pm 10$  & 88 & 2.1 m\\
2451619.81 & 0.601 & $34.4 \pm 0.3$ & $36 \pm 5$   & $-52 \pm 10$  &144 & 2.1 m\\
2451620.80 & 0.528 & $33.6 \pm 0.7$ & $15 \pm 5$   & $89 \pm 10$   & 62 & 2.1 m\\
2451884.82\tablenotemark{a} & 0.785 & $33.9 \pm 0.3$ & $144 \pm 7$  & $-109 \pm 20$ & 71 & 2.7 m\\  
2451885.00\tablenotemark{a} & 0.953 & $34.0 \pm 0.5$ & $45 \pm 12$  & $49 \pm 16$   & 51 & 2.7 m\\  
2451885.81\tablenotemark{a} & 0.712 & $33.9 \pm 0.3$ & $128 \pm 7$  & $-150 \pm 10$ & 72 & 2.7 m\\  
2451886.00\tablenotemark{a} & 0.890 & $33.1 \pm 0.3$ & $107 \pm 7$  & $-29 \pm 10$  & 77 & 2.7 m\\  
2451891.77\tablenotemark{a} & 0.294 & $34.3 \pm 0.5$ & $-68 \pm 12$ & $209 \pm 16$  & 54 & 2.7 m\\
2451891.92\tablenotemark{a} & 0.434 & $33.3 \pm 0.5$ & $0 \pm 12$   & $79 \pm 16$   & 51 & 2.7 m\\
\enddata
\label{rvs}
\tablenotetext{a}{Spectra used in the temperature analysis in \S \ref{tsec}}
\end{deluxetable}

\begin{deluxetable}{llrrr}
\tablewidth{0pc}
\tablecaption{CfA Radial Velocities}
\tablehead{
\colhead{Telescope} & \colhead{HJD} &
\colhead{$v_{\rm rad}$}       & 
\colhead{$\sigma$}            &
\colhead{$o-c$}}
\startdata 
T \dotfill &  2445062.6545 & 31.57 & $ \pm 1.05 $ & $-$3.32 \\
T \dotfill &  2445063.6858 & 35.38 & $ \pm 1.50 $ &    0.50 \\
T \dotfill &  2445064.6825 & 36.62 & $ \pm 1.38 $ &    1.74 \\
T \dotfill &  2445098.6570 & 34.13 & $ \pm 1.46 $ & $-$0.61 \\
M \dotfill &  2445336.9593 & 33.54 & $ \pm 1.01 $ & $-$0.26 \\
M \dotfill &  2445337.9373 & 34.33 & $ \pm 0.67 $ &    0.53 \\
T \dotfill &  2445420.6598 & 30.48 & $ \pm 1.52 $ & $-$2.98 \\
T \dotfill &  2445421.7450 & 33.23 & $ \pm 1.29 $ & $-$0.22 \\
T \dotfill &  2445424.6256 & 32.33 & $ \pm 1.75 $ & $-$1.11 \\
T \dotfill &  2445425.6919 & 37.74 & $ \pm 1.49 $ &    4.30 \\
M \dotfill &  2445475.6292 & 33.93 & $ \pm 0.71 $ &    0.72 \\
M \dotfill &  2445477.6333 & 33.25 & $ \pm 0.93 $ &    0.05 \\
T \dotfill &  2445651.0215 & 33.67 & $ \pm 1.31 $ &    1.40 \\
T \dotfill &  2445652.0279 & 31.35 & $ \pm 1.30 $ & $-$0.92 \\
T \dotfill &  2445654.0270 & 31.40 & $ \pm 1.25 $ & $-$0.86 \\
M \dotfill &  2445708.9448 & 30.44 & $ \pm 1.13 $ & $-$1.44 \\
T \dotfill &  2445721.8978 & 32.15 & $ \pm 1.80 $ &    0.37 \\
T \dotfill &  2445779.6912 & 32.20 & $ \pm 1.11 $ &    0.83 \\
T \dotfill &  2445807.6566 & 31.28 & $ \pm 1.40 $ &    0.05 \\
M \dotfill &  2445831.6563 & 31.52 & $ \pm 0.66 $ &    0.26 \\
T \dotfill &  2445842.6806 & 30.67 & $ \pm 1.20 $ & $-$0.69 \\
T \dotfill &  2446098.8420 & 36.62 & $ \pm 1.06 $ &    1.09 \\
T \dotfill &  2446166.7240 & 34.40 & $ \pm 1.37 $ & $-$0.84 \\
M \dotfill &  2446393.0776 & 34.46 & $ \pm 0.95 $ &    0.14 \\
T \dotfill &  2446421.0072 & 33.23 & $ \pm 1.03 $ & $-$0.99 \\
T \dotfill &  2446429.0448 & 30.18 & $ \pm 1.09 $ & $-$4.00 \\
T \dotfill &  2447496.0633 & 33.03 & $ \pm 0.85 $ & $-$1.63 \\
T \dotfill &  2447515.9366 & 33.95 & $ \pm 0.71 $ & $-$0.63 \\
T \dotfill &  2447544.9637 & 33.51 & $ \pm 1.13 $ & $-$0.96 \\
T \dotfill &  2447554.9214 & 33.37 & $ \pm 1.11 $ & $-$1.06 \\
T \dotfill &  2447573.9151 & 35.57 & $ \pm 1.20 $ &    1.22 \\
T \dotfill &  2447579.8252 & 35.55 & $ \pm 1.00 $ &    1.22 \\
T \dotfill &  2447601.6737 & 34.35 & $ \pm 1.16 $ &    0.10 \\
T \dotfill &  2447602.6281 & 33.37 & $ \pm 0.90 $ & $-$0.87 \\
T \dotfill &  2447603.7199 & 34.79 & $ \pm 1.04 $ &    0.55 \\
T \dotfill &  2447604.7198 & 34.93 & $ \pm 1.08 $ &    0.70 \\
T \dotfill &  2447605.6644 & 33.42 & $ \pm 0.71 $ & $-$0.81 \\
M \dotfill &  2447609.6810 & 35.40 & $ \pm 0.57 $ &    1.18 \\
M \dotfill &  2447610.7042 & 33.80 & $ \pm 0.72 $ & $-$0.41 \\
T \dotfill &  2447611.6206 & 35.36 & $ \pm 1.14 $ &    1.15 \\
T \dotfill &  2447613.7374 & 34.73 & $ \pm 1.41 $ &    0.53 \\
T \dotfill &  2447614.7168 & 35.01 & $ \pm 0.96 $ &    0.81 \\
T \dotfill &  2447629.7150 & 35.69 & $ \pm 1.22 $ &    1.55 \\
T \dotfill &  2447630.6606 & 34.47 & $ \pm 1.17 $ &    0.34 \\
T \dotfill &  2447631.6515 & 33.35 & $ \pm 1.05 $ & $-$0.78 \\
T \dotfill &  2447670.6673 & 34.57 & $ \pm 1.06 $ &    0.59 \\
T \dotfill &  2447841.9893 & 33.97 & $ \pm 0.58 $ &    0.71 \\
T \dotfill &  2447843.9531 & 32.65 & $ \pm 0.65 $ & $-$0.60 \\
T \dotfill &  2447900.8863 & 32.17 & $ \pm 0.82 $ & $-$0.81 \\
T \dotfill &  2447985.6592 & 33.03 & $ \pm 0.81 $ &    0.50 \\
T \dotfill &  2447987.6555 & 32.06 & $ \pm 0.70 $ & $-$0.46 \\
T \dotfill &  2447991.6324 & 31.77 & $ \pm 0.86 $ & $-$0.73 \\
T \dotfill &  2447992.6714 & 32.02 & $ \pm 0.97 $ & $-$0.47 \\
T \dotfill &  2447993.6834 & 32.96 & $ \pm 1.05 $ &    0.47 \\
T \dotfill &  2447994.6567 & 32.81 & $ \pm 0.82 $ &    0.33 \\
T \dotfill &  2447995.6404 & 33.53 & $ \pm 0.67 $ &    1.06 \\
T \dotfill &  2447996.6709 & 32.72 & $ \pm 0.75 $ &    0.25 \\
T \dotfill &  2447997.6818 & 33.70 & $ \pm 0.55 $ &    1.24 \\
T \dotfill &  2448014.6349 & 30.89 & $ \pm 0.91 $ & $-$1.47 \\
T \dotfill &  2448017.6572 & 31.92 & $ \pm 0.78 $ & $-$0.42 \\
T \dotfill &  2448018.6727 & 31.63 & $ \pm 0.74 $ & $-$0.70 \\
T \dotfill &  2448019.6656 & 32.29 & $ \pm 0.89 $ & $-$0.04 \\
T \dotfill &  2448020.6601 & 32.80 & $ \pm 0.81 $ &    0.48 \\
T \dotfill &  2448023.6308 & 32.91 & $ \pm 0.95 $ &    0.61 \\
T \dotfill &  2448024.6270 & 32.10 & $ \pm 0.81 $ & $-$0.20 \\
T \dotfill &  2448025.6291 & 31.91 & $ \pm 0.94 $ & $-$0.38 \\
T \dotfill &  2448257.8928 & 31.76 & $ \pm 0.91 $ & $-$0.70 \\
T \dotfill &  2448284.7919 & 34.25 & $ \pm 0.59 $ &    0.42 \\
T \dotfill &  2448345.7270 & 34.72 & $ \pm 0.70 $ & $-$0.88 \\
T \dotfill &  2448377.6968 & 36.85 & $ \pm 0.78 $ &    1.08 \\
T \dotfill &  2448616.9604 & 34.81 & $ \pm 0.64 $ & $-$0.13 \\
T \dotfill &  2448639.8198 & 33.97 & $ \pm 0.76 $ & $-$0.87 \\
T \dotfill &  2448648.9080 & 36.08 & $ \pm 0.61 $ &    1.28 \\
T \dotfill &  2448672.7514 & 35.31 & $ \pm 0.80 $ &    0.60 \\
T \dotfill &  2448677.8062 & 35.15 & $ \pm 0.79 $ &    0.46 \\
T \dotfill &  2448693.7936 & 33.62 & $ \pm 0.70 $ & $-$1.00 \\
T \dotfill &  2448727.7368 & 33.21 & $ \pm 0.76 $ & $-$1.28 \\
T \dotfill &  2448758.6799 & 33.55 & $ \pm 0.73 $ & $-$0.82 \\
M \dotfill &  2448873.9965 & 32.52 & $ \pm 0.66 $ & $-$1.40 \\
T \dotfill &  2448938.0099 & 32.37 & $ \pm 0.63 $ & $-$1.29 \\
T \dotfill &  2448991.9359 & 33.31 & $ \pm 0.61 $ & $-$0.12 \\
M \dotfill &  2449020.9809 & 33.11 & $ \pm 0.57 $ & $-$0.20 \\
M \dotfill &  2449021.9997 & 34.56 & $ \pm 0.58 $ &    1.26 \\
T \dotfill &  2449030.8067 & 34.29 & $ \pm 0.50 $ &    1.03 \\
T \dotfill &  2449055.7697 & 33.90 & $ \pm 0.94 $ &    0.75 \\
T \dotfill &  2449058.8102 & 32.88 & $ \pm 0.80 $ & $-$0.25 \\
M \dotfill &  2449059.6700 & 34.13 & $ \pm 0.43 $ &    1.00 \\
T \dotfill &  2449078.6871 & 33.99 & $ \pm 0.79 $ &    0.95 \\
M \dotfill &  2449110.6517 & 36.41 & $ \pm 0.73 $ &    3.53 \\
M \dotfill &  2449347.9850 & 33.62 & $ \pm 0.78 $ &    2.27 \\
M \dotfill &  2449351.8640 & 31.97 & $ \pm 0.99 $ &    0.64 \\
T \dotfill &  2449353.9521 & 30.51 & $ \pm 0.87 $ & $-$0.81 \\
M \dotfill &  2449374.7553 & 31.14 & $ \pm 0.96 $ & $-$0.09 \\
T \dotfill &  2449376.9664 & 31.96 & $ \pm 1.01 $ &    0.73 \\
T \dotfill &  2449377.8740 & 31.93 & $ \pm 0.79 $ &    0.71 \\
T \dotfill &  2449406.8289 & 30.05 & $ \pm 0.95 $ & $-$1.29 \\
T \dotfill &  2449465.6614 & 34.01 & $ \pm 0.67 $ &    0.58 \\
M \dotfill &  2449678.0541 & 35.03 & $ \pm 0.62 $ & $-$0.45 \\
T \dotfill &  2449700.9761 & 34.17 & $ \pm 0.66 $ & $-$1.21 \\
M \dotfill &  2449728.8325 & 34.42 & $ \pm 0.69 $ & $-$0.84 \\
T \dotfill &  2449785.8384 & 34.50 & $ \pm 0.59 $ & $-$0.52 \\
T \dotfill &  2449815.6918 & 33.80 & $ \pm 0.86 $ & $-$1.09 \\
T \dotfill &  2450061.9619 & 32.78 & $ \pm 0.71 $ & $-$1.14 \\
T \dotfill &  2450087.8800 & 34.34 & $ \pm 0.56 $ &    0.52 \\
T \dotfill &  2450148.8151 & 33.33 & $ \pm 1.27 $ & $-$0.24 \\
T \dotfill &  2450441.9119 & 29.69 & $ \pm 0.60 $ & $-$2.34 \\
T \dotfill &  2450558.6689 & 29.73 & $ \pm 1.13 $ & $-$1.51 \\
M \dotfill &  2450567.7373 & 32.40 & $ \pm 0.67 $ &    1.18 \\
M \dotfill &  2450798.9724 & 36.91 & $ \pm 0.74 $ &    1.18 \\
T \dotfill &  2450821.8990 & 36.45 & $ \pm 0.79 $ &    0.80 \\
T \dotfill &  2450884.7992 & 36.64 & $ \pm 0.77 $ &    1.24 \\
T \dotfill &  2450914.8058 & 35.08 & $ \pm 1.00 $ & $-$0.19 \\
T \dotfill &  2450920.7580 & 35.91 & $ \pm 0.79 $ &    0.67 \\
T \dotfill &  2450945.7053 & 36.34 & $ \pm 0.83 $ &    1.20 \\
T \dotfill &  2451149.0260 & 35.30 & $ \pm 0.52 $ &    0.98 \\
T \dotfill &  2451176.9691 & 34.88 & $ \pm 0.75 $ &    0.67 \\
T \dotfill &  2451184.9470 & 35.21 & $ \pm 0.63 $ &    1.03 \\
T \dotfill &  2451233.8100 & 34.06 & $ \pm 0.51 $ &    0.07 \\
T \dotfill &  2451243.6833 & 34.03 & $ \pm 0.61 $ &    0.08 \\
T \dotfill &  2451267.7265 & 35.52 & $ \pm 0.83 $ &    1.67 \\
T \dotfill &  2451293.6737 & 34.10 & $ \pm 0.57 $ &    0.35 \\
T \dotfill &  2451533.9451 & 32.22 & $ \pm 0.90 $ & $-$0.41 \\
T \dotfill &  2451566.8947 & 30.90 & $ \pm 0.64 $ & $-$1.54 \\
T \dotfill &  2451595.8628 & 31.64 & $ \pm 0.86 $ & $-$0.62 \\
T \dotfill &  2451619.7617 & 32.03 & $ \pm 0.84 $ & $-$0.07 \\
T \dotfill &  2451623.8223 & 30.97 & $ \pm 0.91 $ & $-$1.11 \\
T \dotfill &  2451650.6877 & 31.77 & $ \pm 0.99 $ & $-$0.12 \\
T \dotfill &  2451835.0027 & 33.84 & $ \pm 0.36 $ &    0.81 \\
T \dotfill &  2451857.0265 & 33.85 & $ \pm 0.98 $ & $-$0.30 \\
T \dotfill &  2451890.0569 & 33.90 & $ \pm 0.81 $ & $-$1.37 \\
T \dotfill &  2451973.7278 & 36.33 & $ \pm 0.75 $ &    0.57 \\
\enddata
\label{cfarv.tab}
\end{deluxetable}

\begin{deluxetable}{lcccccc}
%\hspace*{-0.2in} 
\tablewidth{0pc}
\tablecaption{Component Temperatures and Fluxes from High-Resolution
Spectroscopy}
\tablehead{\colhead{} & \colhead{$T_{Aa}$ (K)} & \colhead{$F_{Aa} / F_{tot}$} & 
\colhead{$T_{Ab}$ (K)} & \colhead{$F_{Ab} / F_{tot}$} & \colhead{$T_{B}$ (K)} & 
\colhead{$F_{B} / F_{tot}$}}
\startdata
Blue Region (456 nm) & $7500 \pm 400$ & $0.33 \pm 0.03$ & $6100 \pm 500$ &
$0.07 \pm 0.03$ & & $0.60$ \\
Red Region (530 nm) & $8300 \pm 250$ & $0.31 \pm 0.03$ & $6100 \pm 400$ &
$0.08 \pm 0.03$ & & $0.61$ \\
Combined & $7775 \pm 450$ & & $6100 \pm 300$ & & $6950 \pm 100$ & \\
Constrained & $7325 \pm 50$ & $0.30 \pm 0.01$ & $6000 \pm 200$ & $0.09 \pm 0.02$ & $6850 \pm 25$ & 0.61 \\
vOVS & $6480\pm25$ & $0.39$ & $5450\pm40$ & $0.19$ & $7500\pm50$ & $0.42$ \\
\enddata
\label{ts}
\end{deluxetable}

\begin{deluxetable}{ccc}
%\hspace*{-0.2in} 
\tablewidth{0pc}
\tablecaption{Light-Curve Solution for Close Binary A}
\tablehead{\colhead{} & \colhead{This Study} & \colhead{vOVS}}
\startdata
$P$ (d) & & $1.0677971 \pm 0.0000007$ \\
$T_{0}$ (d) & & $2444643.250 \pm 0.002$ \\
$q$ & & $0.63 \pm 0.04$ \\
$i$ ($\degr$) & $68 \pm 1$ & $64.0 \pm 1.0$ \\
$T_{Aa}$ (K) & $7325 \pm 50$ \tablenotemark{a} & $6480 \pm 25$
\tablenotemark{b}\\
$T_{Ab}$ (K) & $6000 \pm 200$ \tablenotemark{a} & $5450 \pm 40$
\tablenotemark{b} \\
$T_{Aa} / T_{Ab}$ & $1.22 \pm 0.04$ & $1.19 \pm 0.01$ \\
$R_{Aa} / R_{L}$ & $0.66 \pm 0.03$ & $0.66 \pm 0.07$ \\
$R_{Ab} / R_{L}$ & $0.92 \pm 0.02$ & $0.86 \pm 0.07$ \\ 
$V_{Aa}$ & $12.56 \pm 0.03$ & $12.33 \pm 0.11$\\
$V_{Ab}$ & $13.87 \pm 0.25$ & $13.10 \pm 0.11$\\
$V_{B}$ & $11.79 \pm 0.02$ & $12.24 \pm 0.11$\\
$V_{tot}$ & 11.251\tablenotemark{c} & $11.30 \pm 0.11$\\
$(\bv)_{Aa}$ & $0.33 \pm 0.01$ & $0.51 \pm 0.02$\\
$(\bv)_{Ab}$ & $0.62 \pm 0.07$ & $0.82 \pm 0.02$\\
$(\bv)_{B}$ & $0.428 \pm 0.005$ & $0.22 \pm 0.01$ \\
$(\bv)_{tot}$ & 0.415\tablenotemark{c} & $0.48 \pm 0.02$ \\
\enddata
\label{lccomp}
\tablenotetext{a}{Observational measurement}
\tablenotetext{b}{From light-curve models}
\tablenotetext{c}{Measurements from Montgomery et al. 1993;
used as constraints on temperatures.}
\end{deluxetable}

\begin{deluxetable}{lc}
%\hspace*{-0.2in} 
\tablewidth{0pc}
\tablecaption{CfA Orbit for Narrow-Line Component B}
\tablehead{
\colhead{Parameter} &
\colhead{Value}}
\startdata 
$P$ (days)             \dotfill     & $ 1188.5 \pm 6.76   $ \\
$\gamma$ (km s$^{-1}$) \dotfill     & $  33.76 \pm 0.12   $ \\
$K$ (km s$^{-1}$)      \dotfill     & $   2.28 \pm 0.26   $ \\
$e$                    \dotfill     & $  0.568 \pm 0.076  $ \\
$\omega$ (degrees)     \dotfill     & $    258 \pm 10     $ \\
$T$ (HJD)              \dotfill     & $2448270 \pm 20     $ \\
$a_1 \sin i$ (Gm)      \dotfill     & $   30.7 \pm 6.1    $ \\
$f(M)$ ($M_{\odot}$)   \dotfill     & $0.00081 \pm 0.00048$ \\
$N_{\rm obs}$          \dotfill     &         131           \\
$\sigma$ (km s$^{-1}$) \dotfill     &        1.16           \\
Cycles                 \dotfill     &         5.8           \\
\enddata
\label{cfaorb.tab}
\end{deluxetable}

\begin{deluxetable}{ccc}
\hspace*{-0.2in} 
\tablewidth{0pc}
\tablecaption{Physical Characteristics of S1082 A}
\tablehead{\colhead{} & \colhead{This Study} & \colhead{vOVS}}
\startdata
$T_{Aa}$ (K) & $7325 \pm 50$  & $6480 \pm 25$\\
$T_{Ab}$ (K) & $6000 \pm 200$ & $5450 \pm 40$\\
$T_{B}$ (K) & $6850 \pm 25$ & $7500 \pm 50$ \\
$R_{Aa} (\rsun)$ & $2.04 \pm 0.10$ & $2.07 \pm 0.07$ \\
$R_{Ab} (\rsun)$ & $2.15 \pm 0.10$ & $2.17 \pm 0.07$ \\ 
$M_{Aa} (\msun)$ & $2.52 \pm 0.38$ & $2.70 \pm 0.38$\\
$M_{Ab} (\msun)$ & $1.58 \pm 0.27$ & $1.70 \pm 0.27$\\
$\Omega_{Aa}$ \tablenotemark{a}& $0.73 \pm 0.07$ & $0.49 \pm 0.05$ \\
$\Omega_{Ab}$ \tablenotemark{a}& $0.85 \pm 0.11$ & $0.91 \pm 0.05$ \\
\enddata
\label{phys}
\tablenotetext{a}{Spin axis assumed perpendicular to orbital plane.}
\end{deluxetable}


\begin{thebibliography}{}
\bibitem[Belloni \& Tagliaferri(1998)]{bf98} Belloni, T. \& Tagliaferri, G.
1998, \aap, 335, 517
\bibitem[Belloni \& Verbunt(1996)]{bv96} Belloni, T. \& Verbunt, F.
1996, \aap, 305, 806
\bibitem[Belloni et al.(1998)]{belloni98} Belloni, T., Verbunt, F., \&
Mathieu, R. D. 1998, \aap, 339, 431
\bibitem[Belloni et al.(1993)]{belloni93} Belloni, T., Verbunt, F., \&
Schmitt, J. H. M. M.  1993, \aap, 269, 175
\bibitem[Budding \& Zeilik(1987)]{budd87}Budding, E. \& Zeilik, M. 1987, 
\apj, 319, 827
\bibitem[Carney et al.(2001)]{carn}Carney, B. W., Latham, D. W., Laird, J.
B., Grant, C. E., \& Morse, J. A. 2001, AJ, 122, 3419
\bibitem[Claret \& Cunha(1995)]{clare95}Claret, A.. \& Cunha, N. C. S. 1995, 
\aap, 318, 187
\bibitem[Collier Cameron \& Hilditch(1997)]{coll97}Collier Cameron,
A. \& Hilditch, R. W. 1997, \mnras, 287, 567
\bibitem[Fan et al.(1996)]{fan96}Fan, X. et al. 1996, \aj, 112, 628
\bibitem[Fagerholm(1906)]{fg06}Fagerholm, E. 1906, Ph.D. Thesis, Uppsala Univ.
\bibitem[Girard et al.(1989)]{girard89} Girard, T. M., Grundy, W. M., 
Lopez, C. E., \& Van Altena, W. F. 1989, \aj, 98, 227
\bibitem[Gilliland \& Brown(1988)]{gill88}Gilliland, R. L. \& Brown,
T. M. 1988, \pasp, 100, 754
\bibitem[Glebocki \& Stawikowski(1995)]{gleb95}Glebocki, R. \&
Stawikowski, A. 1995, AcA, 45, 725
\bibitem[Glebocki \& Stawikowski(1997)]{gleb97}Glebocki, R. \&
Stawikowski, A. 1997, \aaps, 328, 579
\bibitem[Goranskij et al.(1992)]{goranskij92}Goranskij, V. P., Kusakin, A. V.,
Mironov, A. V., Moshkaljov, V. G., \& Pastukhova, E. N. 1992, Ast Ap Trans,
2, 201
\bibitem[Gray(1994)]{gray94} Gray, D. F. 1994, \pasp, 106, 1248
\bibitem[Gray \& Johanson(1991)]{gray91} Gray, D. F. \& Johanson, H. L. 1991, 
\pasp, 103, 439
\bibitem[Hauschildt et al.(1999)]{haus99} Hauschildt, P. H., Allard,
F., \& Baron, E. 1999, \apj, 512, 377
\bibitem[Hilditch \& Bell(1994)]{hilditch94}Hilditch, R. W. \& Bell,
S. A. 1994, \mnras, 267, 1081 
\bibitem[Hinkle et al.(2000)]{hinkle00} Hinkle, K., Wallace, L., Valenti, J., 
\& Harmer, D. 2000, Visible and Near Infrared Atlas of the Arcturus Spectrum 
3727-9300 \AA, (San Francisco: ASP)
\bibitem[Honeycutt(1990)]{hone92}Honeycutt, R. K. 1992, \pasp, 104, 435
\bibitem[Hurley et al.(2001)]{hurl01}Hurley, J. R., Tout, C. A., 
Aarseth, S. J., \& Pols, O. R. 2001, \mnras, 323, 630
\bibitem[Kurtz \& Mink(1998)]{KM98}Kurtz, M. J. \& Mink, D. J.
1998, \pasp, 110, 934
\bibitem[Landsman et al.(1997)]{land97}Landsman, W., Aparicio, J.,
Bergeron, P., di Stefano, R., \& Stecher, T. P. 1997, \apjl, 481, L93
\bibitem[Landsman et al.(1998)]{land98}Landsman, W., Bohlin, R. C.,
Neff, Susan, G., O'Connell, R. W., Roberts, M. S., Smith, A. W., \&
Stecher, T. P. 1998, \aj, 116, 789
\bibitem[Latham(1992)]{lath92}Latham, D. W. 1992, in ASP Conf. Ser. 32,
Complementary Approaches to Double and Multiple Star Research, ed.
H. A. McAlister \& W. I. Hartkopf (San Francisco: ASP), 110
\bibitem[Latham \& Milone(1996)]{milo96}Latham, D. W. \& Milone,
A. A. E. 1996, in ASP Conf. Ser. 90, The Origins, Evolutions, \& Destinies
of Binary Stars in Clusters, ed. E. F. Milone \&
J.-C. Mermilliod (San Francisco: ASP), 385
\bibitem[Latham et al.(2002)]{L02}Latham, D. W., Stefanik, R. P.,
Torres, G. T., Davis, R. J., Mazeh, T., Carney, B. W., Laird, J.
B., \& Morse, J. L. 2002, \aj, 124, 1144
\bibitem[Leonard(1996)]{leonard96}Leonard, P. J. T. 1996, \apj, 470, 521
%\bibitem[Leonard \& Linnell(1992)]{leonard92}Leonard, P. J. T. \&
%Linnell, A. P. 1992, \aj, 103, 1928
\bibitem[Manteiga et al.(1989)]{mant89}Manteiga, M., Martinez Roger, C., \&
Pickles, A. J. 1989, \aap, 210, 66
\bibitem[Mardling \& Aarseth(2001)]{mard01}Mardling, R. A. \& Aarseth,
S. J. 2001, \mnras, 321, 398
\bibitem[Mateo et al.(1990)]{mat90}Mateo, M., Harris, H. C., Nemec, J., \&
Olszewski, E. W. 1990, \aj, 100, 469
\bibitem[Mathieu \& Latham(1986)]{mathieu86b}Mathieu, R. D. \& Latham, D. W. 
1986, \aj, 92, 1364
\bibitem[Mathieu et al.(1990)]{mathieu90}Mathieu, R. D., Latham, D. W., \& 
Griffin, R. F. 1990, \aj, 100, 1859
\bibitem[Mathieu et al.(2002)]{mathieu02}Mathieu, R. D., van den Berg,
M., Torres, G., Latham, D. W., Verbunt, F., \& Stassun, K. 2002, AJ,
in press
\bibitem[Mathys(1991)]{mathys91} Mathys, G. 1991, \aap, 245, 467 
 %\bibitem[]{milo91}Milone, A. A. E. 1991, Ph.D. Thesis, Univ. of Cordoba
\bibitem[Milone \& Latham(1992)]{milo92}Milone, A. A. E \& Latham, D. W. 1992,
in IAU Symp. 151, Evolutionary Processes in Interacting Binary Stars, ed. Y.
Kondo, R. F. Sister\'{o}, \& R. S. Polidan (Dordrecht: Kluwer), 475
\bibitem[Milone \& Latham(1994)]{milo94}Milone, A. A. E \& Latham, D. W. 1994,
\aj, 108, 1828
\bibitem[Montgomery et al.(1993)]{montgomery93} Montgomery, K. A., Marschall, 
L. A., \& Janes, K. A. 1993, \aj, 106, 181
\bibitem[Orosz \& Hauschildt(2000)]{oro00}Orosz, J. A. \& Hauschildt, P. H. 2000, \aap, 364, 265
\bibitem[Preston \& Sneden(2000)]{pres}Preston, G. W., \& Sneden, C. 2000,
AJ, 120, 1014
\bibitem[Pribulla et al.(2001)]{prib01}Pribulla, T., Chochol, D.,
Heckert, P. A., Errico, L., Vittone, A. A., Parimucha, S., \&
Teodorani, M. 2001, \aap, 371, 997
\bibitem[Pritchet \& Glaspey(1991)]{pritchet91}Pritchet, C. J., \&
Glaspey, J. W.  1991, \apj, 373, 105
\bibitem[Rappaport et al.(1995)]{rapp}Rappaport, S., Podsiadlowski, P., Joss,
P. C., di Stefano, R., \& Han, Z. 1995, MNRAS, 273, 731
%\bibitem[Sandage(1953)]{sandage53}Sandage, A. R. 1953, \apj, 58, 61
\bibitem[Sanders(1977)]{sanders77} Sanders, W. L. 1977, \aaps, 27, 89
\bibitem[Sandquist \& Shetrone(2002)]{sand02} Sandquist, E. L. \&
Shetrone, M. D. 2002, \aj, submitted
\bibitem[Schaller et al.(1992)]{schall92}Schaller, G., Schaerer, D.,
Meynet, G., \& Maeder, A. 1992, \aaps, 96, 269
\bibitem[Shetrone \& Sandquist(2000)]{shet00} Shetrone, M. D. \&
Sandquist, E. L. 2000, \aj, 120, 1913
\bibitem[Sills et al.(2001)]{sills01}Sills, A., Faber, J. A.,
Lombardi, J. C., Jr., Rasio, F. A., \& Warren, A. R. 2001, \apj, 548,
323
\bibitem[Simoda(1991)]{sim91}Simoda, M. 1991, IBVS, 3657, 1
\bibitem[Singh et al.(1996)]{singh96}Singh, K. P., Drake, S. A., \&
White, N. E. 1996, \aj, 111, 2415
\bibitem[Soderblom et al.(1993)]{soderblom93} Soderblom, D. R.,
Stauffer, J. R., Hudon, J. D., \& Jones, B. F. 1993, \apjs, 85, 315
\bibitem[Stetson(1990)]{stet90}Stetson, P. B. 1990, \pasp, 102, 932
\bibitem[Strassmeier(2000)]{strass00}Strassmeier, K. G. 2000, \aap, 357, 608
\bibitem[Tassoul \& Tassoul(1992)]{tass92} Tassoul, J.-L. \& Tassoul,
M. 1992, \apj, 395, 259
\bibitem[Tull et al.(1995)]{tull95} Tull, R. G., MacQueen, P. J., Sneden, C., 
\& Lambert, D. L., 1995, \pasp, 107, 251
\bibitem[van den Berg et al.(2001)]{vdb01}van den Berg, M., Orosz, J.,
Verbunt, F., \& Stassun, K. 2001, \aap, 375, 375 (vOVS)
\bibitem[van den Berg et al.(1999)]{vdb99}van den Berg, M., Verbunt, F., 
\& Mathieu, R. D. 1999, \aap, 347, 866
\bibitem[Vogt et al.(1999)]{vogt99}Vogt, S. S., Hatzes, A. P., Misch, A. A., 
\& K\"{u}rster, M. 1999, \apjs, 121, 547
\bibitem[Zhao et al.(1993)]{zhao93} Zhao, J. L., Tian, K. P., Pan, R. S., 
He, Y. P., \& Shi, H. M. 1993, \aaps, 100, 243
\end{thebibliography}
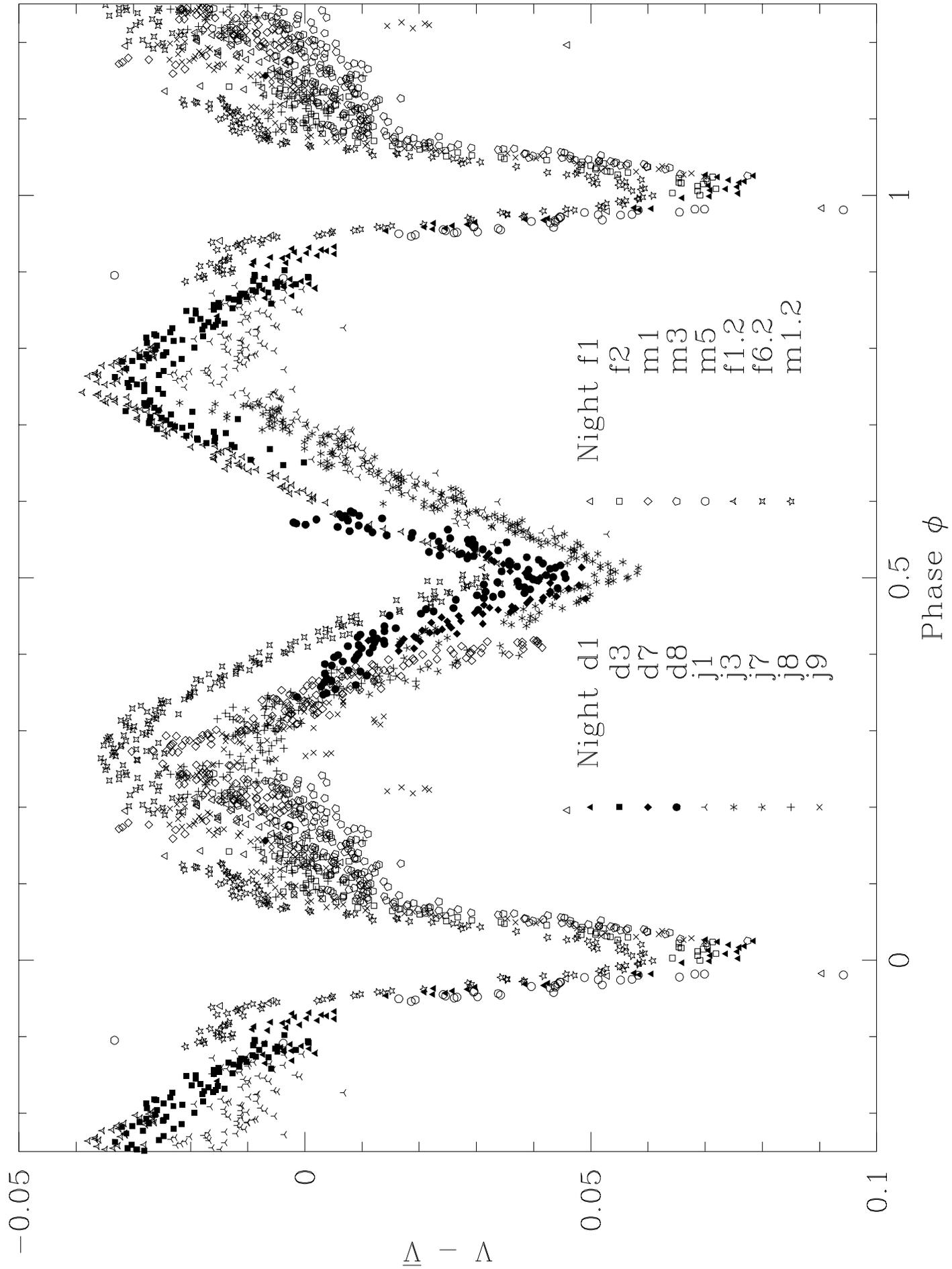
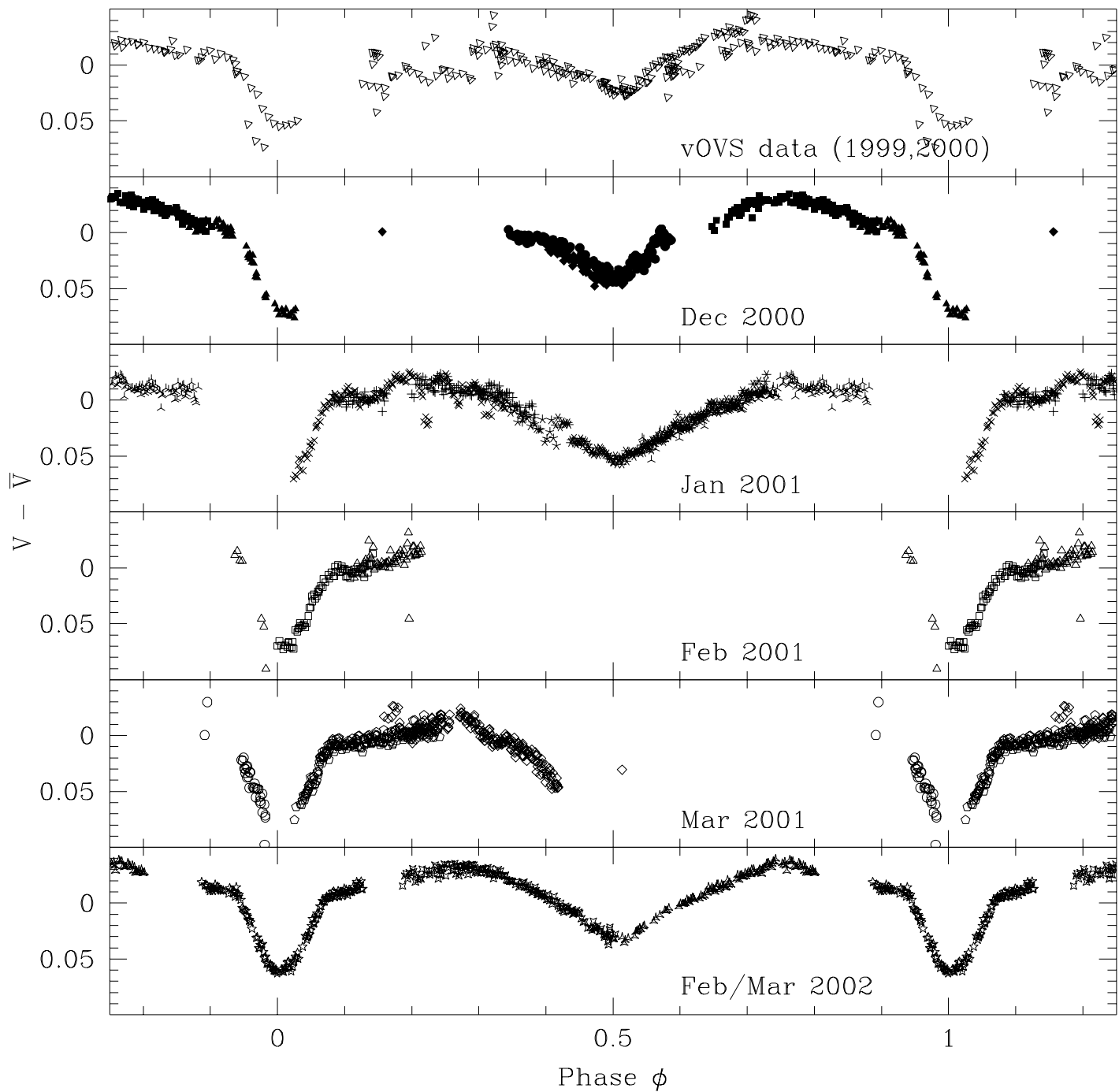
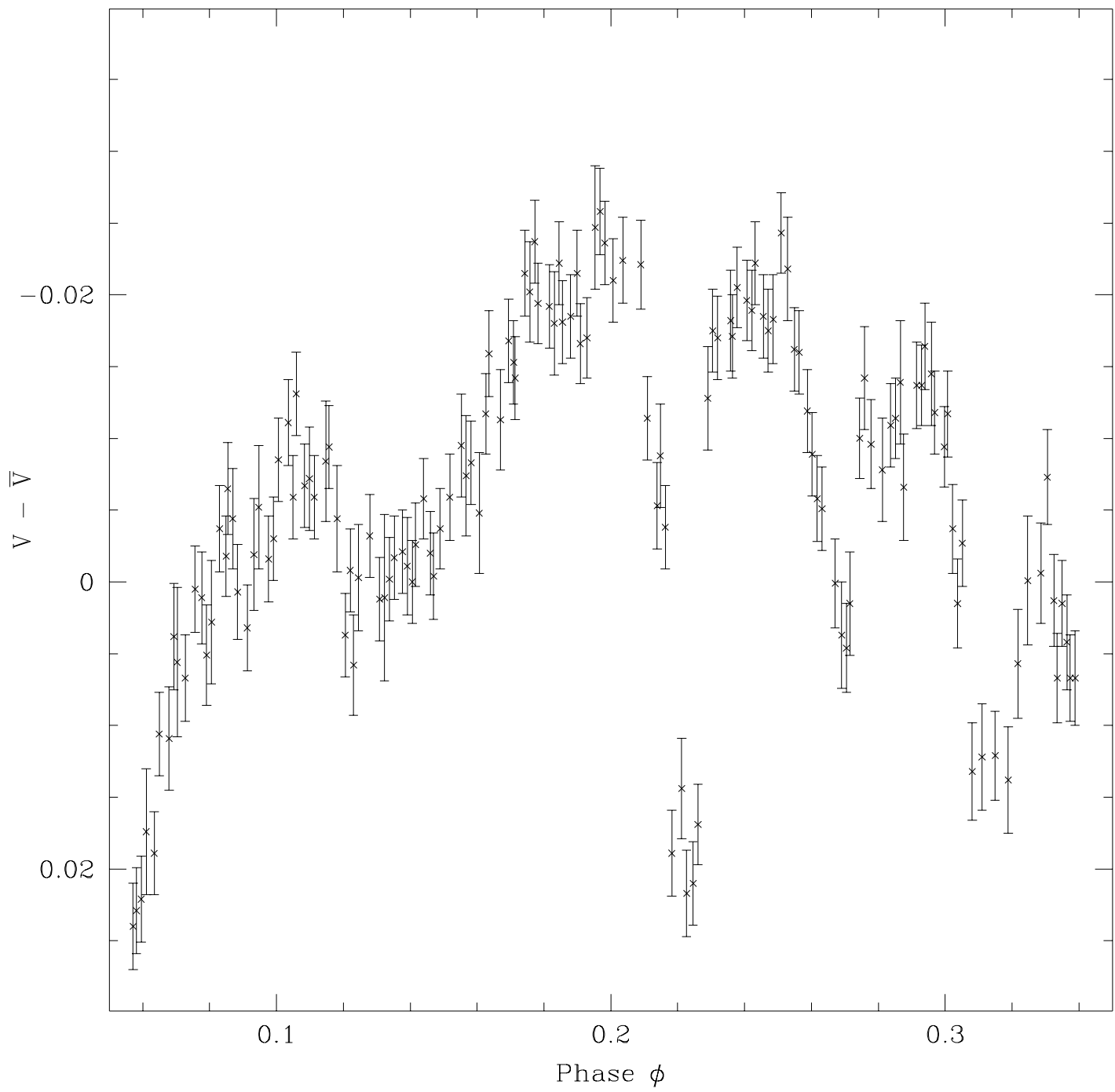
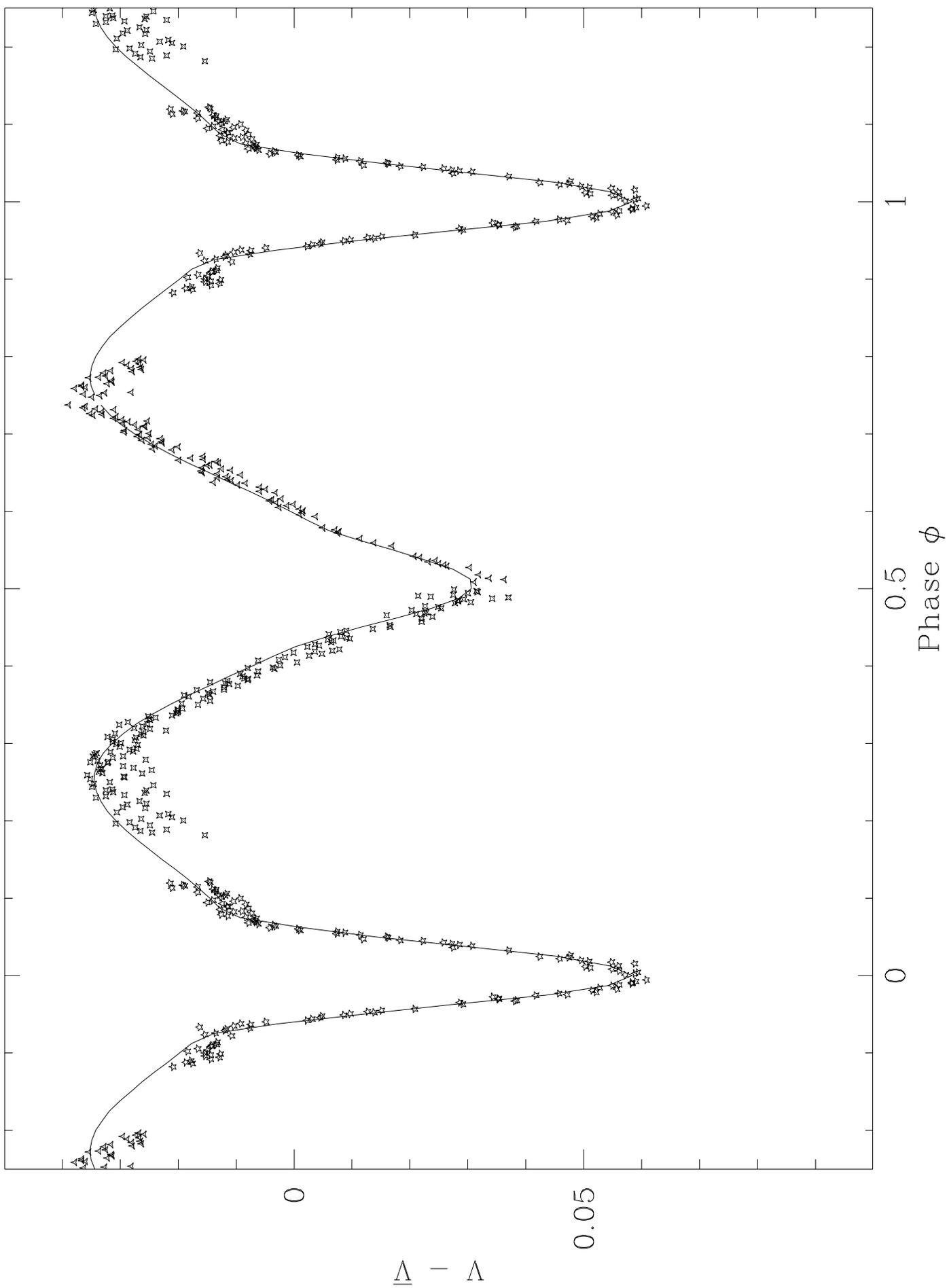
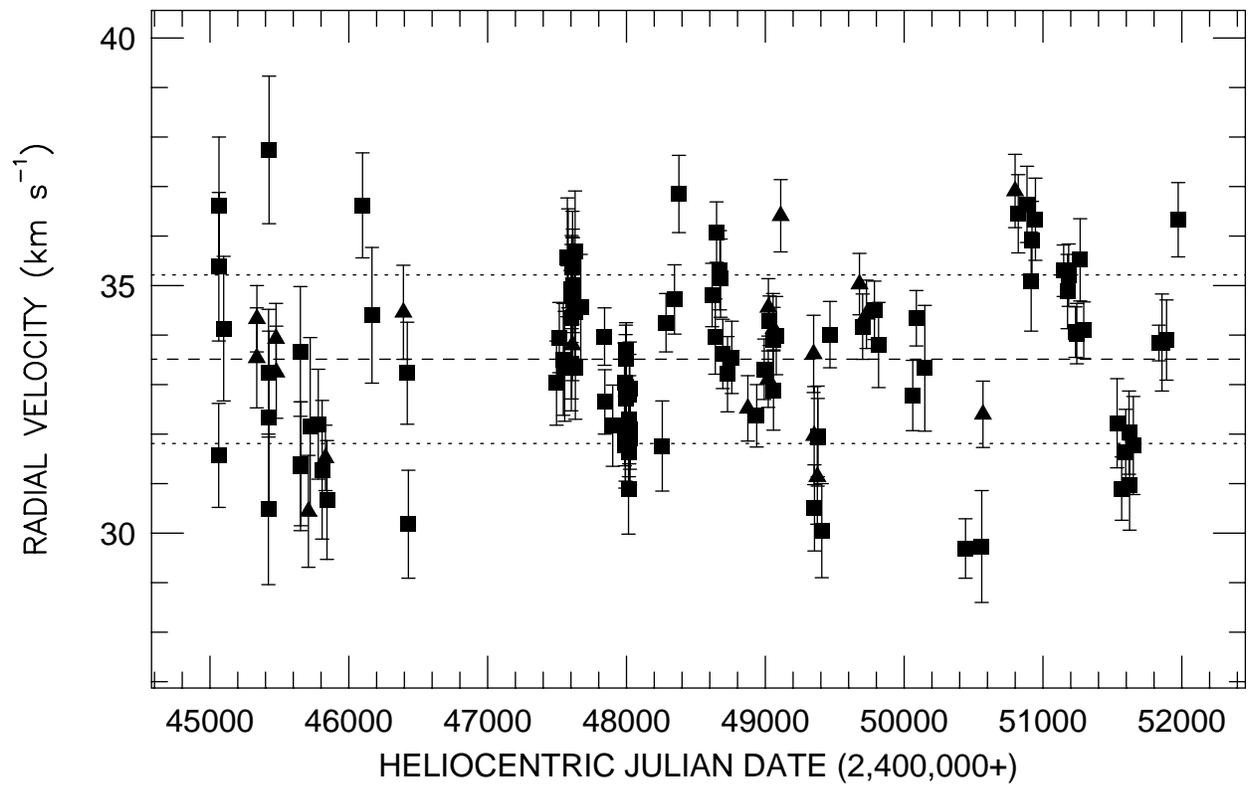
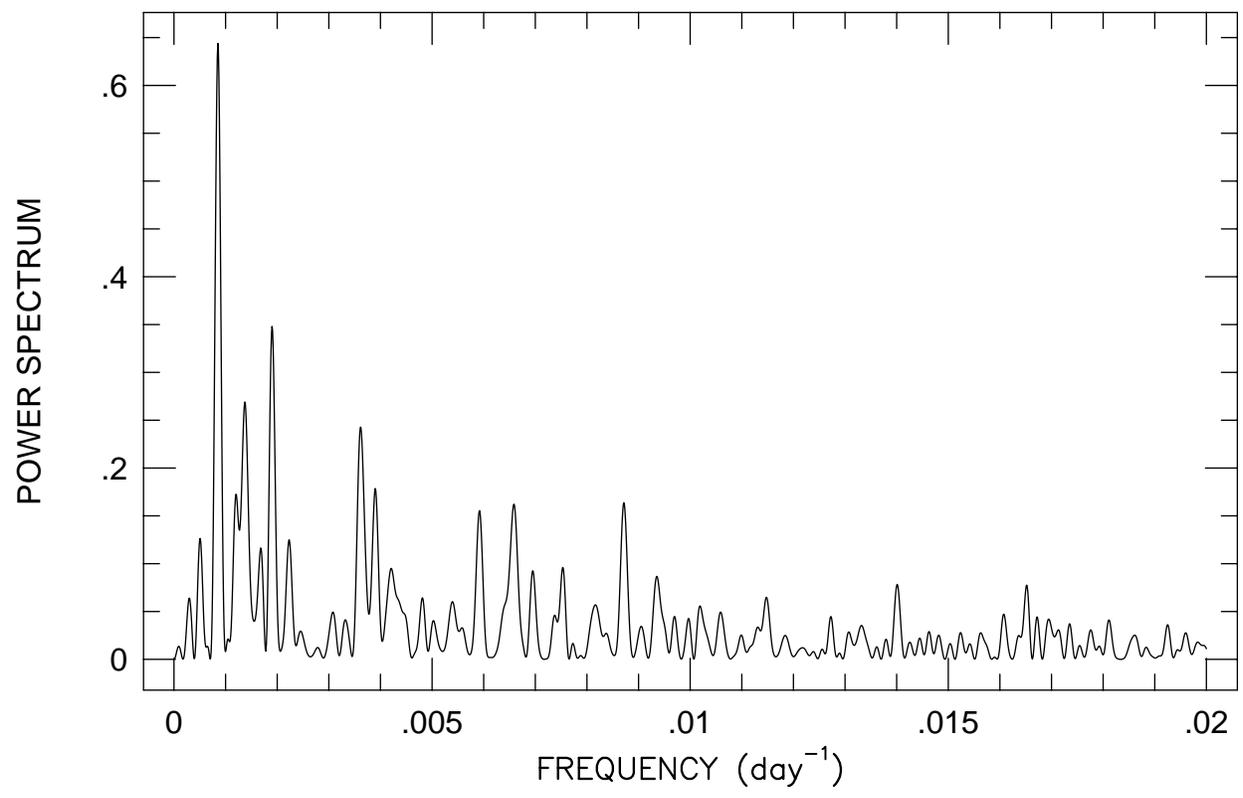
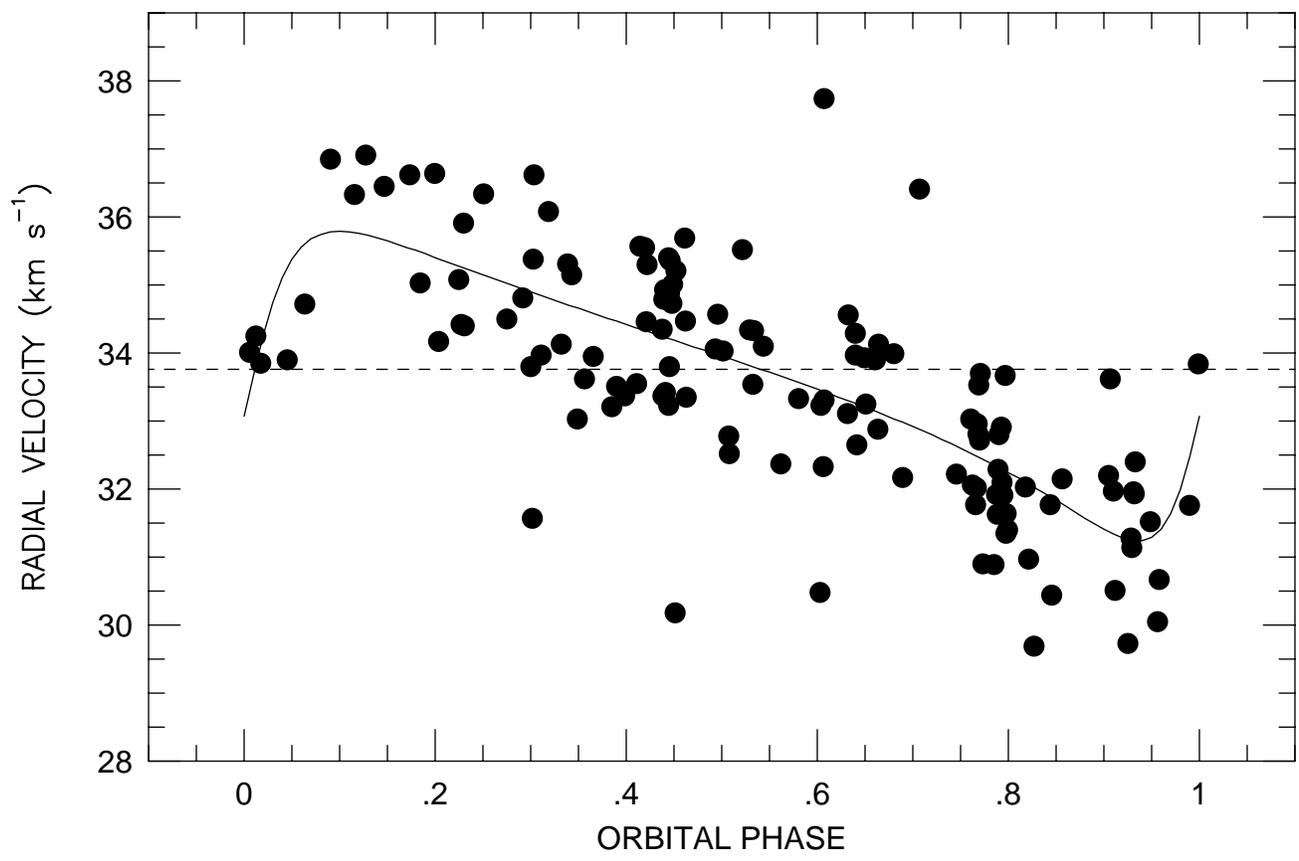
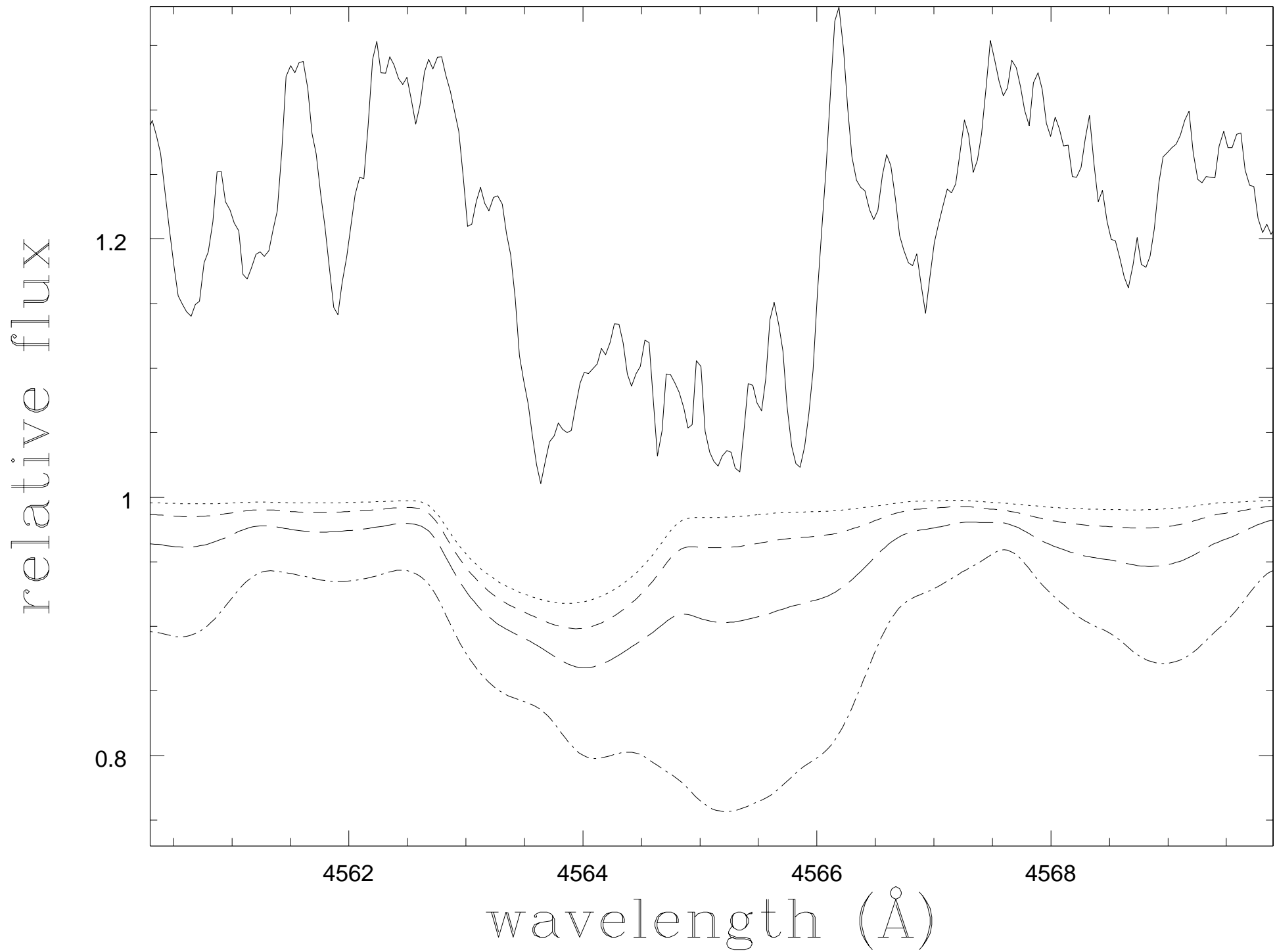
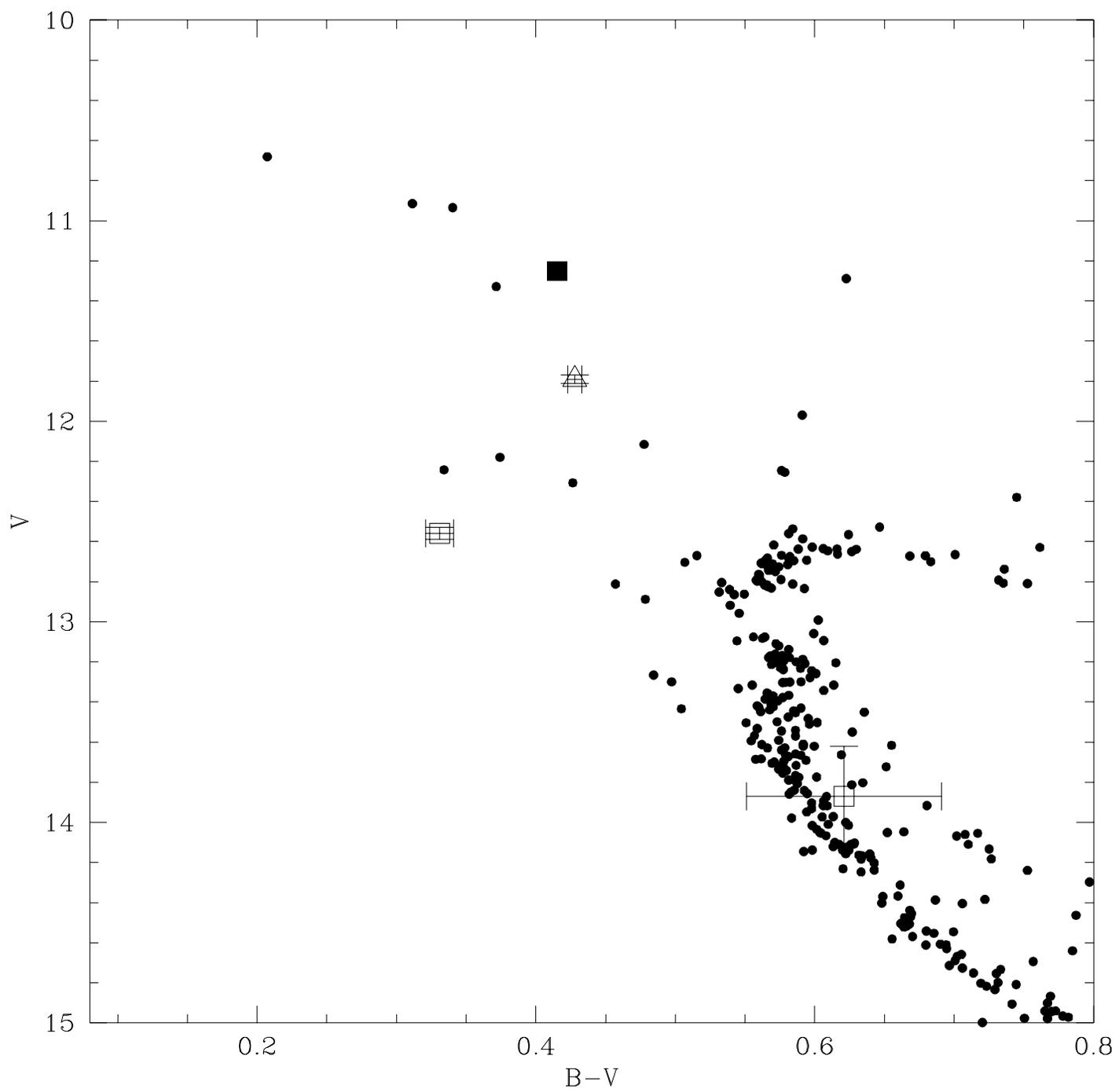
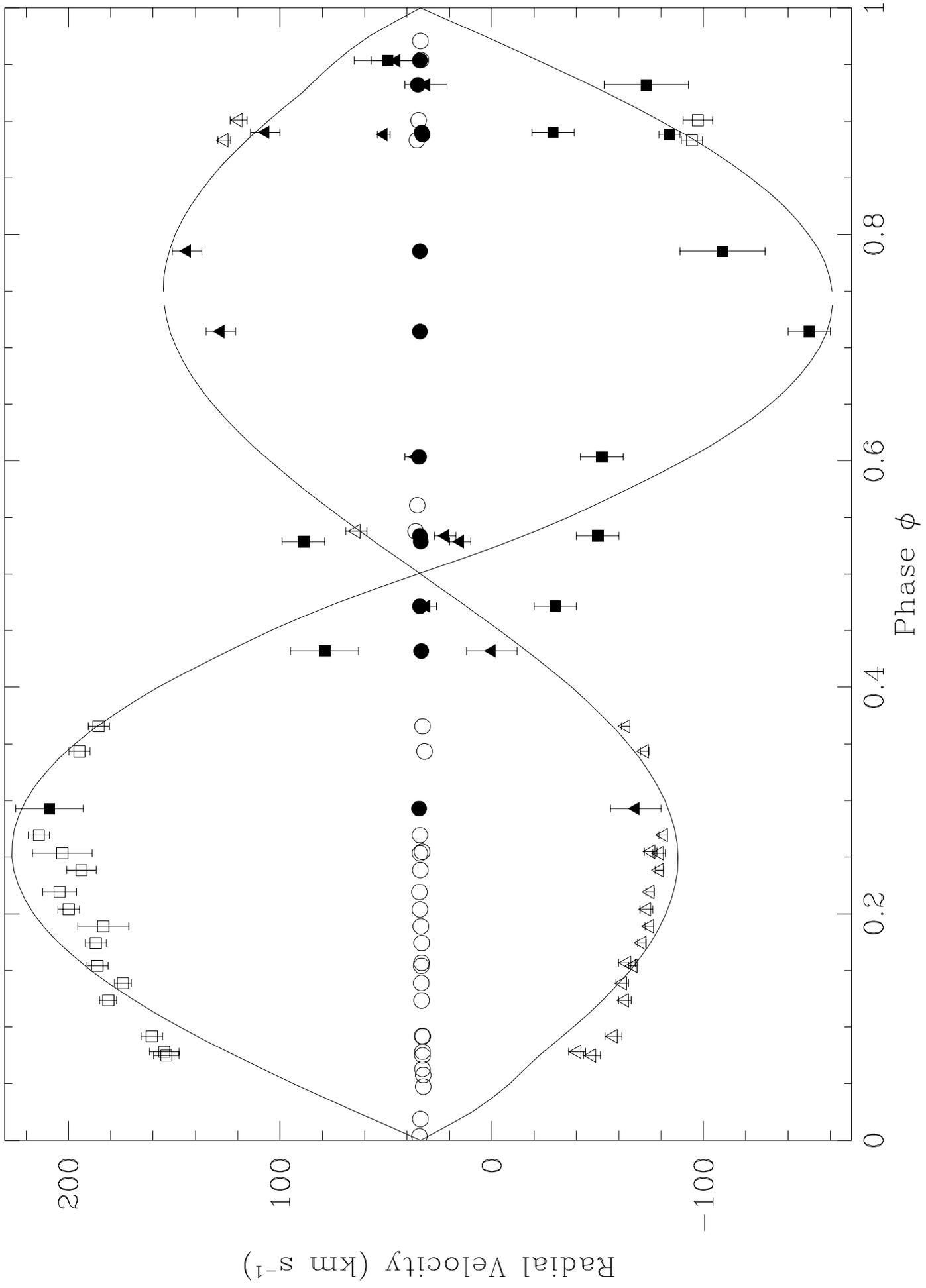
\end{document}